\documentclass[12pt]{article}
\usepackage{graphicx}
\usepackage{amsmath}
\usepackage{amssymb}
\usepackage{ulem}
\usepackage{enumerate}
\usepackage{cite}
\usepackage{latexsym}
\begin{document}
\begin{center}
{\bf {\large{Abelian 2-Form Gauge Theory: Basic Canonical Brackets and Nilpotency Property of the Noether (Anti-)BRST Charges }}}

\vskip 3.0cm

{\sf  R. P. Malik$^{(a,b)}$}\\
$^{(a)}$ {\it Physics Department, Institute of Science,}\\
{\it Banaras Hindu University, Varanasi - 221 005, India}\\

\vskip 0.1cm

$^{(b)}$ {\it DST Centre for Interdisciplinary Mathematical Sciences,}\\
{\it Institute of Science, Banaras Hindu University, Varanasi - 221 005, India}\\
{\small {\sf {e-mails: rpmalik1995@gmail.com; malik@bhu.ac.in}}}
\end{center}

\vskip 2.0 cm

\noindent
{\bf Abstract:} Within the framework of Becchi-Rouet-Stora-Tyutin (BRST) formalism, we invoke the beauty of the
{\it basic} canonical (anti)commutators to prove the nilpotency property of the Noether (anti-)BRST charges for 
the BRST-quantized D-dimensional free Abelian 2-form gauge theory which
is endowed with a {\it non-trivial} Curci-Ferrari (CF) type restriction. In this proof, we use {\it only} the 
theoretical strength of the partial integration along with the
Gauss divergence theorem. We demonstrate that, under the off-shell nilpotent (anti-)BRST transformations, the
Noether conserved (anti-)BRST charges are {\it not} invariant and they are also found to be {\it not} off-shell nilpotent (if
we exploit the standard relationship between the continuous symmetry transformations and their generators as the Noether conserved charges). However,
these charges become (anti-)BRST invariant and nilpotent  if we use (i) the appropriate equations of motion
at suitable places, and (ii) the partial integration along with the Gauss divergence theorem. We derive the consistently {\it modified}
versions of the Noether (anti-)BRST charges which are {\it invariant} under the off-shell nilpotent (anti-) BRST transformations. We 
discuss the physicality criteria w.r.t. (i) the conserved Noether (anti-)BRST charges, and (ii) the modified (anti-)BRST invariant 
versions of the Noether (anti-)BRST charges. We prove the superiority of the {\it latter} over the {\it former} (in view of the consistency with the Dirac
quantization conditions for the gauge theories).

\vskip 0.8cm
\noindent
PACS numbers: 11.15.-q; 12.20.-m; 03.70.+k \\

\vskip 0.5cm
\noindent
{\it {Keywords}}: BRST-quantized D-dimensional Abelian 2-form gauge theory;  off-shell nilpotent (anti-)BRST transformations; 
Noether theorem and conserved charges; basic canonical (anti)commutators; nilpotency property of (anti-)BRST charges; physicality criteria

\newpage

\section {Introduction}

The study of the higher $p$-form ($p = 2, 3, 4... $) gauge theories has become an interesting area of research because 
of the fact that the {\it basic} 
$p$-form ($p = 2, 3, 4... $) fields appear in the quantum excitations of the (super)string theories (see, e.g. [1-5] and references therein) which
are the forefront areas of research activities in 
the domain of theoretical high energy physics (THEP). The reach and range of the (super)string theories
are expected to go beyond the realm of the standard model of elementary particle physics (SMEPP)
which is one of the most successful theories in THEP
because of the stunning degree of agreements between {\it its} theoretical predictions and experimental verifications. 
However, there are a few serious drawbacks\footnote{The SMEPP provides a {\it unified} theoretical
description of the three (i.e. electromagnetism, weak and strong) 
out of the total {\it four} fundamental interactions of nature. However, it does {\it not} say
anything about the theoretical aspects of gravitation at the 
classical as well as at the quantum level. In addition, the conclusive experimental evidence of the masses of
neutrinos is one of the biggest blow to the very foundation of the  basic theoretical ideas behind the 
existence of the SMEPP. This observation demonstrates that the SMEPP is {\it not} a complete theory 
in the true sense of the word and one has to go beyond it. The (super)string theories, at present, are the 
most promising candidates to do so. These theories are, in fact, expected to
provide (i) the {\it unified} theoretical description of {\it all} the four fundamental interactions of nature, (ii) the consistent theory of
quantum gravity, (iii) the theoretical aspects of SMEPP as the low energy limit, etc.} in the
theoretical aspects of SMEPP which are expected to be resolved
by the (super)string theories. The modern theoretical developments in the realm
of the {\it latter} theories have influenced the research activities in the domain of 
pure mathematics, too.  As a consequence, there has been 
a very interesting confluence and convergence of ideas from theoretical aspects of (super)strings and some of the key concepts of pure mathematics.
In this context, it is pertinent to point out that
we have devoted quite a  number of years
to study {\it especially} higher Abelian $p$-form (i.e. $p = 2, 3 $) gauge theories by exploiting the theoretical strength of the 
Becchi-Rouet-Stora-Tyutin (BRST) formalism [6-9]. Particularly, we have been able to establish that the BRST-quantized Abelian
$p$-form ($p = 1, 2, 3 $)  {\it massless} (see, e.g. [10-13] and references therein) and
the St{\" u}ckelberg-modified {\it massive} gauge theories (see, e.g. [14-16] and references therein) are the examples for the
tractable field-theoretic
models of Hodge theory in $D = 2\, p $ (i.e. D = 2,4,6) dimensions of the Minkowskian spacetime. In such kinds of research activities,
there has been convergence of ideas from (i) the physical aspects of BRST formalism, and (ii) the mathematical ingredients that are 
associated with the de Rham cohomological operators of differential geometry 
(see, e.g., [17-22] and references therein) at the {\it algebraic} level.

The main motivating factor that has propelled our keen interest in pursuing our present investigation is as follows. In our very recent work
on the BRST-quantized version of the D-dimensional non-Abelian 1-form gauge theory [23],
we have applied the theoretical strength of the {\it basic}  canonical (anti)commutators to prove (i) the off-shell nilpotency of the Noether conserved 
(anti-)BRST charges where {\it only} the sanctity of the 
partial integration along with the Gauss divergence theorem has played an important role,  and (ii) the (anti-)BRST invariance of the 
{\it modified} versions of the  Noether conserved (anti-)BRST charges where $(a)$ the appropriate equations of motion
at suitable places, and $(b)$ the
partial integration along with the theoretical strength of Gauss's
divergence theorem have been taken into account.  
It is a well-known fact that the BRST-quantized version of the non-Abelian 1-form gauge theory is
endowed with a  non-trivial Curci-Ferrari (CF) condition [24]. To corroborate the results of 
our previous work [23], we have taken up the
BRST-quantized version of the D-dimensional 
free Abelian 2-form gauge theory which is {\it also} endowed with a non-trivial CF-type restriction
(see, e.g. [15,11] and references therein). We are sure that our present investigation will put the results of (i) our {\it previous} work [23]
on the BRST-quantized version of the D-dimensional non-Abelian 1-form,  and (ii) {\it present} work on the BRST-quantized 
version of the D-dimensional Abelian 2-form gauge theories, on the
firmer footings of theoretical physics (that are connected with the BRST formalism).

Against the backdrop of the above paragraph, we would like to further elaborate on the reasons behind our
interest in the present investigation. 
First, we have demonstrated, in our earlier work [23],
that the (anti-)BRST invariant {\it modified} versions of the conserved Noether (anti-)BRST charges play a very important role in the
physicality criteria (within the framework of BRST formalism) because they lead to the conditions on the physical {\it quantum}  states 
that are found to be consistent with the Dirac quantization conditions (for theories that are endowed with any kinds of constraints). Second,
the nilpotency property of the Noether (anti-)BRST charges are {\it true} if and only if we utilize the validity 
of the Gauss divergence theorem when we use the basic canonical (anti)commutators in this proof [23]. On the other hand,
when the relationship between the continuous symmetry transformations and {\it their} generators as the Noether conserved charges 
is taken into consideration, we have found that (i) the appropriate Euler-Lagrange (EL) equations of motion (EoM) at suitable places, 
and (ii) the Gauss divergence theorem, 
{\it both} are needed in the proof of the nilpotency property of the Noether (anti-)BRST charges. 
Third, it turns out
that the (anti-)BRST invariance of the {\it modified} versions of the Noether 
(anti-)BRST charges are true if we utilize (i) the basic canonical
(anti)commutators, and (ii) the EL-EoM along with the validity of Gauss's divergence theorem. Fourth, using the basic canonical brackets, we have shown that
the {\it modified} versions of the Noether (anti-)BRST charges are {\it not} nilpotent
{\it unless} we use the EL-EoM along with the Gauss divergence theorem at appropriate places. Finally, we have concluded that (i) the Noether conserved
(anti-)BRST charges, and (ii) the  consistently  {\it modified} 
(anti-)BRST invariant versions of the Noether (anti-)BRST charges, have their own importance
and identity (within the framework of the BRST  formalism).

The theoretical materials of our present investigation are organized as follows. In the next section, we (i) recapitulate the bare essentials of 
the off-shell
nilpotent (anti-)BRST transformations for the coupled (but equivalent) Lagrangian densities, and (ii) derive the Noether conserved (anti-)BRST charges 
for our BRST-quantized theory. Our Sec. 3 deals with (i) the derivations, and
(ii) the utility  of the
{\it basic} (anti)commutators, to prove that the Noether (anti-)BRST charges are the generators for the {\it off-shell} nilpotent (anti-)BRST
transformations. We {\it also} comment clearly on the (anti-)BRST 
invariance of {\it these} charges. The subject matter of our Sec. 4 is concerned with the systematic derivations of the (anti-) BRST invariant
versions of the (anti-)BRST charges where we have taken into account (i) the partial integration along with Gauss's divergence theorem, 
(ii) the appropriate
EL-EoM, and (iii) the off-shell nilpotent (anti-)BRST symmetry 
transformations {\it together}. Our Sec. 5 contains the theoretical content that is connected with the discussion on the 
physicality criteria w.r.t. the {\it above} (i) Noether (anti-)BRST charges, and (ii) the {\it modified} versions 
of the (anti-)BRST charges.  In Sec. 6, we devote time on the discussion of the nilpotency property of the conserved (anti-)BRST charges
from different theoretical angles.
Finally, in our Sec. 7, we summarize our key
results, mention a few novel observations and point out the future perspective of our present investigation.

In our Appendices A, B, C, D and E, we perform some explicit computations that are useful in clarifying a few statements that have been made in the
main body of our text.\\

{\it Conventions and Notations:} We choose the flat D-dimensional background  Minkowskian spacetime
metric tensor  $\eta_{\mu\nu} $ = diag (+1, -1, -1, -1, .....) so that the dot
product between two non-null vectors $U_\mu$ and $V_\mu$ is defined as : $U \cdot V = \eta_{\mu\nu} U^\mu V^\nu = U_0 V_0 - U_i V_i $ 
where (i) the Greek indices $\mu, \nu, \sigma... = 0, 1, 2...(D-1) $ correspond to the time and space directions, 
(ii) the Latin indices $i, j, k...= 1, 2, 3...(D-1) $ stand for the {\it space} directions {\it only}, and (iii) the repeated indices are 
{\it always}  summed over.
 In the entire body of our text,
we adopt the convention of the left-derivative w.r.t. the fermionic fields of our theory at {\it appropriate} places. We follow 
the convention: $(\delta B_{\mu\nu}/\delta B_{\rho\sigma}) = \frac{1}{2!}\, (\delta^\rho_\mu \delta^\sigma_\nu - \delta^\rho_\nu \delta^\sigma_\mu) $
for the variation/differentiation of the antisymmetric tensor (i.e. 2-form) gauge field $B_{\mu\nu}$.
The symbols used to denote the
off-shell nilpotent (i.e. $s_{(a)b}^2 = 0$) (anti-)BRST transformation operators are $s_{(a)b} $. On the other hand, 
we always represent the {\it Noether} (anti-)BRST  
charges by the notations $Q_{(a)b} $. We freely use the brackets [...] and (...) in our entire text in  
a {\it normal} fashion. However,
in our equations, we have denoted the (anti)commutator(s) by the notations $\{...\}$ and [...], respectively. The over dot (i.e. $\dot \Phi $)
 on the generic field $\Phi$
of our theory corresponds to the time derivative [i.e. $\dot \Phi = \partial_0 \Phi \equiv (\partial \Phi/\partial t) $] in the
natural units where $\hbar = c = 1 $.


\section{Preliminaries: Coupled Lagrangian Densities}

In our earlier works on the 4D BRST-quantized version of (i) the {\it massless} free Abelian 2-form gauge theory [11], and (ii) 
the St{\" u}ckelberg-modified {\it massive} Abelian 2-form (i.e. $p = 2 $) gauge theory [15], we have 
systematically derived the coupled (but equivalent) Lagrangian densities. In our present investigation, we shall focus on the {\it massless}
versions of the coupled Lagrangian densities that have been obtained in our earlier work [15]
which are {\it true} for our present\footnote{We very briefly sketch (cf. Appendix A)
the derivations of the gauge-fixing and Faddeev-Popov (FP) ghost terms 
of the coupled (but equivalent) Lagrangian  densities ${\cal L}_{(B)} $ and ${\cal L}_{(\bar B)} $ of our BRST-quantized theory
by taking into account the off-shell nilpotent (anti-)BRST symmetry transformation operators $s_{(a)b}$.}  
BRST-quantized version of the D-dimensional free Abelian
2-form gauge theory, too. These {\it perfectly} BRST and anti-BRST invariant 
Lagrangian  densities ${\cal L}_{(B)} $ and ${\cal L}_{(\bar B)} $, respectively, are as follows (see, e.g. [15,11]) 
\begin{eqnarray}\label{1}
{\cal L}_{(B)} &=& \dfrac{1}{12}\, H^{\mu\nu\sigma}\,H_{\mu\nu\sigma} + B^\mu \Big [\partial^\nu B_{\nu\mu} - \dfrac{1}{2} \, \partial_\mu \phi \Big ]
- \dfrac{1}{2} \, B^\mu\, B_\mu - \Big (\partial_\mu \bar C_\nu - \partial_\nu \bar C_\mu \Big )\, \partial^\mu C^\nu \nonumber\\
 &-& \dfrac{1}{2}\,
\partial_\mu \bar \beta \,\partial^\mu \beta 
- \dfrac{1}{2} \, \Big [(\partial \cdot C) - \frac{1}{4} \, \lambda \Big ] \, \rho 
- \dfrac{1}{2} \, \Big [(\partial \cdot \bar C) + \frac{1}{4} \, \rho \Big ]\, \lambda, \nonumber\\
{\cal L}_{(\bar B)} &=& \dfrac{1}{12}\, H^{\mu\nu\sigma}\,H_{\mu\nu\sigma} 
- \bar B^\mu \Big [\partial^\nu B_{\nu\mu} + \dfrac{1}{2} \, \partial_\mu \phi \Big ]
- \dfrac{1}{2} \, \bar B^\mu\, \bar B_\mu - \Big (\partial_\mu \bar C_\nu - \partial_\nu \bar C_\mu \Big )\, \partial^\mu C^\nu \nonumber\\
 &-& \dfrac{1}{2}\,
\partial_\mu \bar \beta \,\partial^\mu \beta 
- \dfrac{1}{2} \, \Big [(\partial \cdot C) - \frac{1}{4} \, \lambda \Big ] \, \rho 
- \dfrac{1}{2} \, \Big [(\partial \cdot \bar C) + \frac{1}{4} \, \rho \Big ]\, \lambda,
\end{eqnarray}
where the kinetic term (i.e. $ \frac{1}{12}\, H^{\mu\nu\sigma}\,H_{\mu\nu\sigma} $) for the 
Abelian 2-form [i.e. $B^{(2)} = \frac{1}{2!}\, B_{\mu\nu}\, (d x^\mu \wedge d x^\nu) $] antisymmetric tensor gauge field $B_{\mu\nu}$
owes its origin to the Abelian 3-form [i.e. $H^{(3)} = d\, B^{2} \equiv \frac{1}{3!}\, H_{\mu\nu\sigma}\, (d x^\mu \wedge d x^\nu \wedge dx^\sigma) $].
Here the field-strength tensor $H_{\mu\nu\sigma} = \partial_\mu B_{\nu\sigma} + \partial_\nu B_{\sigma\mu} + \partial_\sigma B_{\mu\nu} $
is totally antisymmetric in {\it all} its indices because we note:  $d = \partial_\mu \; (dx^\mu) $  [with $d^2 = \frac{1}{2!}\, (\partial_\mu \partial_\nu
- \partial_\nu \partial_\mu) \, (dx^\mu \wedge dx^\nu ) = 0 $] is the extrior derivatve of differerntial geometry [17-22].
These Lagrangian densities are called as the {\it coupled}
Lagrangian densities because the {\it bosonic} Nakanishi-Lautrup auxiliary fields $B_\mu$ and  $\bar B_\mu$  are
{\it not} independent. To be precise, these fields are restricted to satisfy the CF-type restriction:
$B_\mu + \bar B_\mu + \partial_\mu \phi = 0$
 where the scalar field $\phi$ has been introduced into our theory because of the stage-one reducibility that is 
associated with the free Abelian 2-form antisymmetric 
(i.e. $B_{\mu\nu} = -\, B_{\nu\mu}$) tensor gauge field $B_{\mu\nu}$. In the above equation (1), to be precise,
 we have invoked the bosonic Nakanishi-Lautrup auxiliary
fields to linearize the gauge-fixing terms for the gauge field $B_{\mu\nu}$ in two different (i.e. linearly independent) ways. It is
straightforward to check that the CF-type restriction is satisfied (cf. Appendix B for details)
if we use the Euler-Lagrange (EL) equations of motion (EoM) w.r.t. the
Nakanishi-Lautrup auxiliary fields $B_\mu$ and  $\bar B_\mu$ that are present in the non-ghost sectors
of the {\it coupled} Lagrangian densities [cf. Eq. (1)] of our D-dimensional BRST-quantized version of the free Abelian 2-form gauge theory.

As far as the Faddeev-Popov (FP) ghost part of the Lagrangian  density is concerned, we note that
the fermionic (i.e. $C_{\mu}^{2}=\bar{C}_{\mu}^{2}=0, C_{\mu} \bar{C}_{\nu}+\bar{C}_{\nu} C_{\mu}=0, C_{\mu} C_{\nu}+C_{\nu} C_{\mu}=$ $0, 
\bar{C}_{\mu} \bar{C}_{\nu}+\bar{C}_{\nu} \bar{C}_{\mu}=0$) Lorentz vector (anti-)ghost fields $(\bar C_\mu)C_\mu$ carry the
ghost numbers $(-1)+1$ and  the (anti-)ghost fields $(\bar{\beta}) \beta$ are  the Lorentz scalar 
{\it bosonic}  (i.e. $\beta^2 = 0, \; \bar \beta^2 = 0, \; \beta\, \bar \beta = \bar \beta \, \beta$)
(anti-)ghost fields (i.e. the ghost-for-ghost fields) are endowed with the 
ghost numbers $(-2)+2$, respectively . The 
fermionic (i.e. $\rho^2 = \lambda^2 = 0, \; \rho\,\lambda  + \lambda\, \rho = 0 $) auxiliary (anti-) ghost fields:
$\rho = -\, 2\, (\partial \cdot \bar C), \;  \lambda = +\, 2\, (\partial \cdot  C)$ carry the ghost numbers (-1)+1, respectively. These
(anti-)ghost fields are invoked in our BRST-quantized theory  to maintain the sanctity of the {\it unitarity} 
at any arbitrary order of perturbative computation for any physical process that is 
theoretically/experimentally  {\it allowed} by our present BRST-quantized theory.

It is straightforward to check that, under the following 
infinitesimal, continuous and off-shell nilpotent (i.e. $s_{(a)b}^2 = 0 $)
(anti-)BRST symmetry transformations ($s_{(a)b}$), namely;
\begin{eqnarray}\label{2}
&& s_{ab} B_{\mu\nu} = -\, (\partial_\mu \bar C_\nu - \partial_\nu \bar C_\mu), \quad s_{ab} \bar C_\mu = -\, \partial_\mu \bar \beta, 
\quad s_{ab}  C_\mu = \bar B_\mu, 
\quad s_{ab} \phi = +\, \rho, \nonumber\\
&&s_{ab}  \beta = -\, \lambda, \qquad \;s_{ab}  B_\mu = -\, \partial_\mu \rho, \qquad \;
s_{ab} \big [\lambda, \;\rho, \; \bar \beta,  \; \bar B_\mu, \; H_{\mu\nu\sigma} \big ] = 0, \nonumber\\
&& s_b B_{\mu\nu} = -\, (\partial_\mu C_\nu - \partial_\nu C_\mu), \qquad s_b C_\mu = -\, \partial_\mu \beta, \qquad s_b \bar C_\mu = B_\mu, 
\qquad s_b \phi = \lambda, \nonumber\\
&&s_b \bar \beta = -\, \rho, \qquad \; s_b \bar B_\mu = -\, \partial_\mu \lambda, \qquad \;
s_b \big [\lambda, \;\rho, \;\beta, \; B_\mu, \; H_{\mu\nu\sigma} \big ] = 0,
\end{eqnarray}
the Lagrangian  densities ${\cal L}_{(\bar B)}$ and ${\cal L}_{(B)}$ transform to the {\it total}
 spacetime derivatives as
\begin{eqnarray}\label{3}
s_{ab} {\cal L}_{(\bar B)} &=& +\, \partial_\mu \Big [ (\partial^\mu \bar C^\nu - \partial^\nu \bar C^\mu)\, \bar B_\nu 
+ \dfrac{1}{2}\, \lambda\, \partial^\mu \bar \beta - 
\dfrac{1}{2}\, \rho\, \bar B^\mu \Big ], \nonumber\\
s_b {\cal L}_{(B)} &=& -\, \partial_\mu \Big [ (\partial^\mu C^\nu - \partial^\nu C^\mu)\, B_\nu + \dfrac{1}{2}\, \lambda\, B^\mu - 
\dfrac{1}{2}\, \rho\, \partial^\mu\, \beta \Big ], 
\end{eqnarray}
thereby rendering the action integrals  $S_1 = \int d^D x \,{\cal L}_{(\bar B)} $ and $S_2 = \int d^D x \,{\cal L}_{(B)} $
invariant (i.e. $ s_{ab} S_1 = 0 , \; s_b S_2 = 0$) for the well-defined physical fields that vanish off as $x \to \pm\, \infty $ due to the validity of the
celebrated Gauss divergence theorem.

According to Noether's theorem, the invariance of the action integrals (e.g. $ s_{ab} S_1 = 0, \; s_b S_2 = 0 $) under the infinitesimal, continuous 
and off-shell nilpotent (anti-)BRST symmetry transformations (2) leads to the derivations of the conserved Noether currents which, in turn, lead to the
derivations of the Noether conserved (anti-)BRST charges. These exercises have been thoroughly 
performed  in our earlier work on the St{\" u}ckelberg-modified massive Abelian 2-form gauge theory [15] whose {\it massless}
limit obviously yield the Noether conserved (anti-)BRST charges [for the sake of our present discussion(s)].
We quote here the explicit expressions for the (anti-)BRST Noether conserved currents:
\begin{eqnarray}\label{4}
J^\mu_{ab} &=& \big (\partial^\mu C^\nu - \partial^\nu C^\mu \big ) \, \partial_\nu \bar \beta + \big (\partial^\mu \bar C^\nu 
- \partial^\nu \bar C^\mu \big ) \, \bar B_\nu  \nonumber\\
&+& \frac{1}{2}\, \lambda \, \partial^\mu \bar \beta - \frac{1}{2}\, \rho \, \bar B^\mu
- \dfrac{1}{2}\, H^{\mu\nu\sigma} \,
(\partial_\nu \bar C_\sigma - \partial_\sigma \bar C_\nu) , \nonumber\\
J^\mu_{b} &=& - \Big [\big (\partial^\mu \bar C^\nu - \partial^\nu \bar C^\mu \big ) \, \partial_\nu  \beta + \big (\partial^\mu  C^\nu 
- \partial^\nu  C^\mu \big ) \,  B_\nu \nonumber\\
&-& \frac{1}{2}\, \rho \, \partial^\mu  \beta + \frac{1}{2}\, \lambda \,  B^\mu  +  \dfrac{1}{2}\, H^{\mu\nu\sigma} \,
(\partial_\nu C_\sigma - \partial_\sigma C_\nu) \Big ].
\end{eqnarray}
The conservation law (i.e. $\partial_\mu J^\mu_{(a)b} = 0 $) can be proven by taking into account the appropriate EL-EoM that are derived from the Lagrangian  
densities ${\cal L}_{(\bar B)}$ and ${\cal L}_{(B)}$. The explicit expressions for the Noether conserved (anti-)BRST charges  
i.e. $Q_{(a)b} = \int d^{D-1} x\, J^{0}_{(a)b} $) for our {\it massless} 
D-dimensional BRST-quantized free Abelian 2-form gauge theory are: 
\begin{eqnarray}\label{5}
Q_{ab} &=& \int d^{D-1} x\,\, \Big [\big (\partial^0 C^i - \partial^i C^0 \big ) \, \partial_i \bar \beta + \big (\partial^0 \bar C^i 
- \partial^i \bar C^0 \big ) \, \bar B_i \nonumber\\
&+& \frac{1}{2}\, \lambda \, \dot {\bar \beta} - \frac{1}{2}\, \rho \, \bar B^0 - \dfrac{1}{2}\, H^{0ij} \,
(\partial_i \bar C_j - \partial_j \bar C_i)\Big ], \nonumber\\
Q_{b} &=& - \int d^{D-1} x\,\,
\Big [\big (\partial^0 \bar C^i - \partial^i \bar C^0 \big ) \, \partial_i  \beta + \big (\partial^0  C^i 
- \partial^i  C^0\big ) \,  B_i \nonumber\\
&-& \frac{1}{2}\, \rho \, \dot  \beta + \frac{1}{2}\, \lambda \,  B^0 + \dfrac{1}{2}\, H^{0ij} \,
(\partial_i C_j - \partial_j  C_i) \Big ].
\end{eqnarray}
We shall see, in our next section, that {\it these} Noether conserved (anti-)BRST charges are found to be the generators for the 
infinitesimal, continuous and off-shell nilpotent (anti-) BRST 
transformations (2), from which, they are derived by using the Noether theorem.

We conclude our present section  with the following crucial remarks. First of all, we note that the field-strength tensor $H_{\mu\nu\sigma} $
(for the Abelian 2-form gauge field $B_{\mu\nu}$) remains invariant under the off-shell nilpotent (anti-)BRST symmetry transformations (2). 
It is worthwhile to recall that the field-strength tensor  $H_{\mu\nu\sigma} $ owes its origin to the exterior derivative of differential geometry
because we have seen that the Abelian 3-form $H^{(3)} =  d\, B^{(2)} $ leads to the definition/derivation of  $H_{\mu\nu\sigma} $. Second,
the nilpotent (anti-)BRST symmetry transformations are {\it fermionic} in nature. Thus, they transform the bosonic field into fermionic field and
vice-versa. Third, the absolute anticommutativity property (i.e. $\{ s_b, \; s_{ab} \} = 0$) between the off-shell nilpotent 
BRST and anti-BRST symmetry operators
{\it distinguishes} them from a couple of ${\mathcal N} = 2 $ SUSY transformation 
operators which are nilpotent of order two {\it but} they do not anticommute with each-other. The {\it latter} property is very 
sacrosanct in the context of the ${\mathcal N} = 2 $ SUSY transformation operators. Fourth, the (anti-)BRST invariant CF-type restriction
(i.e. $B_\mu + \bar B_\mu + \partial_\mu \phi = 0$) is responsible for (i) the absolute anticommutativity property (i.e. $\{ s_b, \; s_{ab} \} = 0$) 
between the BRST and
anti-BRST symmetry transformation operators $s_{(a)b} $, and (ii) the existence of the coupled (but equivalent)  Lagrangian  
densities ${\cal L}_{(\bar B)}$ and ${\cal L}_{(B)}$  [cf. Eq. (B.5) below]. Finally, we note that the Noether 
conserved (anti-)BRST charges $Q_{(a)b} $ are {\it not} invariant under the {\it off-shell}
nilpotent (i.e. $s_{(a)b}^2 = 0 $)
and absolutely anticommuting (i.e. $\{ s_b, \;  s_{ab} \} = 0 $)
(anti-)BRST transformation operators (2) because we observe that the following expressions are true, namely;
\begin{eqnarray}\label{6}
s_{ab} Q_{ab} &=& \int d^{D-1} x\, \Big [\big (\partial^0 \bar B^i - \partial^i \bar B^0 \big ) \, \partial_i \bar \beta \Big ] \neq 0, \nonumber\\
s_b Q_{b} &=& - \int d^{D-1} x\, \Big [\big (\partial^0  B^i - \partial^i  B^0 \big ) \, \partial_i  \beta \Big ] \neq 0.
\end{eqnarray}
Hence, these Noether conserved (anti-)BRST charges $Q_{(a)b} $ are {\it not} physical in the true sense of the word (within the framework of BRST formalism).
Thus, we have to find out the consistently {\it modified} versions of 
the Noether conserved charges $Q_{(a)b} $ which remain {\it invariant} under the infinitesimal, continuous
and {\it off-shell} nilpotent (anti-)BRST transformations (2). This requirement is {\it important} when we discuss the physicality criteria 
(cf. Sec. 5 below)
w.r.t. the (anti-)BRST charges within the framework of BRST formalism. We would like to remark that if we use the appropriate EL-EoM of our theory,
the integrands of equation (6) can become total {\it space} derivatives. At this stage, if we exploit the 
theoretical strength of  the Gauss divergence theorem,
the Noether (anti-)BRST charges become invariant (cf. Sec. 6 for more discussions). However, since we are dealing with 
the {\it off-shell} nilpotent (anti-)BRST symmetry transformations,
we are {\it not} allowed to use the EL-EoM of our D-dimensional BRST-quantized free Abelian 2-form gauge theory. \\


\section{Basic Canonical (Anti)commutators: Application}

Our present section is divided into two parts. In our Subsec. 3.1, we obtain the covariant canonical quantization of our
D-dimensional BRST-quantized version of the free Abelian 2-form gauge theory. Our Subsec. 3.2 is devoted to the proof that
the Noether conserved (anti-)BRST charges [cf. Eq. (5)] are the generators for the off-shell nilpotent (anti-)BRST symmetry
transformations (2) if we use the {\it basic} canonical brackets of our theory.\\


\subsection{Covariant Canonical Quantization}

One of the key features of any arbitrary gauge theory is the observation that it is described by a singular Lagrangian density
which respects a gauge symmetry transformation. As a consequence, 
such kinds of theories (based on the gauge symmetries) 
are {\it always} endowed with the first-class constraints in the terminology of Dirac's prescription
for the classification scheme of constraints [25-31]. For instance, in the case of our present gauge theory, we note that
there are {\it two} fist-class constraints on our theory at the {\it classical} level (cf. Appendix C) where the primary constraint is such
that one of the components of the canonical conjugate momenta w.r.t. the gauge field
turns out to be zero. Hence, we face the problem of the {\it covariant} 
canonical quantization of such kinds of theories. One of the highlights of the BRST-quantization scheme, however, is the observation
that such kinds of issues do {\it not} arise at all
(when we quantize a given gauge theory using the BRST formalism). We discuss, very concisely, this {\it fact} here in the next paragraph.

For the covariant canonical quantization of our D-dimensional field-theoretic system of the free Abelian 2-form gauge theory, first of
all, we have to derive the expressions for the canonical conjugate momenta (i.e. $\Pi $'s)
w.r.t. the {\it dynamical} fields of the coupled (but equivalent)
Lagrangian densities ${\cal L}_{(\bar B)}$ and ${\cal L}_{(B)}$ [cf. Eq. (1)] of our BRST-quantized theory. Toward this goal in mind, let 
us focus on the {\it dynamical} fields: $B_{\mu\nu}, \, \bar C_\mu, \, C_\mu, \, \bar \beta, \beta, \phi $.
The canonical conjugate momenta w.r.t. these dynamical fields are:
\begin{eqnarray}\label{7}
 \Pi_{(B)}^{\mu\nu} (b) &=& \dfrac{\partial \, {\cal L}_{(B)}}{\partial\, (\partial_0\, B_{\mu\nu})} = \dfrac{1}{2}\, \Big [H^{0\mu\nu}
+ \big (\eta^{0\mu} \, B^\nu -  \eta^{0\nu} \, B^\mu \big ) \Big ] \nonumber\\
&\;\;\Longrightarrow\;\;& \Pi_{(B)}^{0i} (b) = \dfrac{1}{2} \, B^i \equiv -\,\dfrac{1}{2} \, B_i , \quad  \Pi_{(B)}^{ij} (b) = \dfrac{1}{2}\, H^{0ij}, \nonumber\\
 \Pi_{(B)}^{\mu\nu} (\bar b) &=& \dfrac{\partial \, {\cal L}_{(\bar B)}}{\partial\, (\partial_0\, B_{\mu\nu})} = \dfrac{1}{2}\, \Big [H^{0\mu\nu}
- \big (\eta^{0\mu} \, \bar B^\nu -  \eta^{0\nu} \, \bar B^\mu \big ) \Big ] \nonumber\\
&\;\;\Longrightarrow\;\;& \Pi_{(B)}^{0i} (\bar b) = -\, \dfrac{1}{2} \, \bar B^i \equiv +\,\dfrac{1}{2} \, \bar B_i , \qquad 
 \Pi_{(B)}^{ij} (\bar b) = \dfrac{1}{2}\, H^{0ij}, \nonumber\\
\Pi^\mu_{(C)}  &=& \dfrac{\partial \, {\cal L}_{(B, \bar B)}}{\partial\, (\partial_0\, C_\mu)} = \Big (\partial^0 \bar C^\mu - \partial^\mu \bar C^0 \Big )
- \dfrac{1}{2} \, \eta^{0\mu}\, \rho, \nonumber\\
&\;\;\Longrightarrow\;\;& \Pi^0_{(C)} = - \dfrac{1}{2} \; \rho, \qquad \Pi^i_{(C)} = \Big (\partial^0 \bar C^i - \partial^i \bar C^0 \Big ) \nonumber\\
\Pi^\mu_{(\bar C)}  &=& \dfrac{\partial \, {\cal L}_{(B, \bar B)}}{\partial\, (\partial_0\, \bar C_\mu)} = -\,\Big (\partial^0  C^\mu - \partial^\mu  C^0 \Big )
- \dfrac{1}{2} \, \eta^{0\mu}\, \lambda \nonumber\\
&\;\;\Longrightarrow\;\;& \Pi^0_{(\bar C)} = - \dfrac{1}{2} \; \lambda, \qquad 
\Pi^i_{(\bar C)} = -\, \Big (\partial^0  C^i - \partial^i  C^0 \Big ), \nonumber\\
\Pi_{(\beta)}  &=& \dfrac{\partial \, {\cal L}_{(B, \bar B)}}{\partial\, (\partial_0\, \beta)} = -\,\dfrac{1}{2}\, \dot {\bar \beta}, \qquad
\Pi_{(\bar \beta)}  = \dfrac{\partial \, {\cal L}_{(B, \bar B)}}{\partial\, (\partial_0\, \bar \beta)} = -\,\dfrac{1}{2}\, \dot {\beta}, \nonumber\\
 \Pi_{(\phi)} (b) &=& \dfrac{\partial \, {\cal L}_{(B)}}{\partial\, (\partial_0\, \phi)} = -\, \dfrac{1}{2}\, B^0, \qquad
 \Pi_{(\phi)} (\bar b) = \dfrac{\partial \, {\cal L}_{(\bar B)}}{\partial\, (\partial_0\, \phi)} = -\, \dfrac{1}{2}\, \bar B^0, 
\end{eqnarray} 
where (i) the parenthesis $(\bar b)$ and $(b)$ in front of some of the above 
canonical conjugate momenta denote the fact that they have been derived from the 
Lagrangian densities ${\cal L}_{(\bar B)}$ and ${\cal L}_{(B)}$, respectively, 
(ii) the subscripts, associated with the above conjugate momenta, are the shorthand notations 
for the dynamical fields: $B_{\mu\nu}, \, \bar C_\mu, \, C_\mu, \, \bar \beta, \beta, \phi $,
and (iii) the symbol $ \eta^{\mu\nu}$ is the inverse (i.e. $\eta_{\mu\nu} \, \eta^{\nu\sigma} = \delta_\mu^\sigma \equiv \eta^{\sigma\nu} \, \eta_{\nu\mu} $)
of the metric tensor $\eta_{\mu\nu} $ that has been chosen for our D-dimensional flat Minkowskian spacetime.
Since the FP-ghost parts of the 
Lagrangian densities ${\cal L}_{(\bar B)}$ and ${\cal L}_{(B)}$ [cf. Eq. (1)] are the {\it same} for our theory, we have 
{\it not} incorporated any kinds of 
parenthesis in front of the  canonical conjugate momenta (i.e. $\Pi$'s) that are associated with the 
{\it dynamical} fermionic/bosonic (anti-)ghost fields. It is worthwhile to mention here that, within the framework of 
BRST formalism, {\it all} the components of the fields (that are present in the {\it coupled} Lagrangian 
densities ${\cal L}_{(\bar B)}$ and ${\cal L}_{(B)}$ [cf. Eq. (1)]) have their corresponding canonical conjugate momenta. As a consequence, there is 
{\it no} problem (of any kinds) as far as the basic ingredients of the covariant canonical quantization scheme is concerned. This is why
the BRST-quantization scheme is one of the most beautiful approaches to the quantization of gauge theories (which are at the heart of the
theoretical description of the {\it three} out of four fundamental interactions of nature).

At this juncture, we are in the position to covariantly quantize our theory in the sense that we can define 
the non-vanishing canonical (anti)commutators for the D-dimensional BRST-quantized version of the
free Abelian 2-form gauge theory that is described by the {\it coupled} 
Lagrangian densities ${\cal L}_{(\bar B)}$ and ${\cal L}_{(B)}$ [cf. Eq. (1)]. The following general {\it basic} canonical brackets are {\it true}
for {\it both} these coupled Lagrangian densities, namely; 
\begin{eqnarray}\label{8}
&&{\left[\phi (\vec{x}, t), \; \Pi_{(\phi)} (\vec{y}, t)\right] }  = i \,\delta^{(D-1)}(\vec{x}-\vec{y}),
\quad {\left[\beta (\vec{x}, t), \; \Pi_{(\beta)} (\vec{y}, t)\right] }  = i \,\delta^{(D-1)}(\vec{x}-\vec{y}),
\nonumber\\
&& {\left[\bar \beta (\vec{x}, t), \; \Pi_{(\bar \beta)} (\vec{y}, t)\right] }  = i \,\delta^{(D-1)}(\vec{x}-\vec{y}), \quad
{\left \{C_0 (\vec{x}, t), \; \Pi^0_{(C)} (\vec{y}, t)\right \} }  = i \,\delta^{(D-1)}(\vec{x}-\vec{y}), \nonumber\\
&& {\left \{ {\bar C}_0 (\vec{x}, t), \; \Pi^0_{({\bar C})} (\vec{y}, t)\right \} }  = i \,\delta^{(D-1)}(\vec{x}-\vec{y}), \;
{\left \{ {C}_i (\vec{x}, t), \; \Pi^j_{({C})} (\vec{y}, t)\right \} }  = i \,\delta_i^j\, \delta^{(D-1)}(\vec{x}-\vec{y}),
\nonumber\\
&&{\left[\bar C_{i}(\vec{x}, t), \; \Pi_{(\bar C)}^{j}(\vec{y}, t)\right] }  = i \,\delta_{i}^{j} \,\delta^{(D-1)}(\vec{x}-\vec{y}),\;\;
{\left[B_{0 i}(\vec{x}, t), \; \Pi_{(B)}^{0 j}(\vec{y}, t)\right] }  = \frac{i}{2!}\,\delta_{i}^{j} \,\delta^{(D-1)}(\vec{x}-\vec{y}), \nonumber\\
&&{\left[B_{i j}(\vec{x}, t), \Pi_{(B)}^{k l}(\vec{y}, t)\right] }  =\frac{i}{2!}\,
\left(\delta_{i}^{k} \delta_{j}^{l}-\delta_{i}^{l} \delta_{j}^{k}\right) \,\delta^{(D-1)}(\vec{x}-\vec{y}),
\end{eqnarray}
where (i) the above brackets (i.e. commutators and anticommutators) are known as the equal-time {\it basic} canonical (anti)commutators,
and (ii) we have taken into account the natural units where $\hbar=c=1$ in {\it their}
definitions [cf. Eq. (8)]. All the rest of the equal-time canonical (anti)commutators, as per the {\it rules} of the canonical quantization scheme, 
are equal to {\it zero} for our D-dimensional BRST-quantized free Abelian 2-form gauge theory. \\


\subsection{Noether Conserved (Anti-)BRST Charges as Generators}

The central purpose of our present subsection is to utilize the basic canonical brackets (8) to prove that the Noether conserved 
(anti-)BRST charges [cf. Eq. (5)] are the generators for the off-shell nilpotent (anti-)BRST symmetry transformations (2). Toward this goal
in mind, first of all, we express the Noether conserved (anti-)BRST charges [cf. Eq. (5)] 
in terms of the canonical conjugate momenta [cf. Eq. (7)], namely;
\begin{eqnarray}\label{9}
 Q_{ab} &=& \int d^{D-1} x\, \Big [ 2\, \Pi^0_{(\bar C)}\, \Pi_{(\beta)} + 2\, \Pi^i_{(C)}\, \Pi^{0i}_{(B)} (\bar b) - \Pi^i_{(\bar C)} \, \partial_i \bar \beta
 \nonumber\\
 &-& \Pi^{ij}_{(B)} (\bar b) \, \big(\partial_i \bar C_j - \partial_j \bar C_i \big ) - 2\, \Pi^0_{(C)}\, \Pi_{(\phi)} (\bar b) \Big ] (\vec{x}, t), \nonumber\\
Q_{b} &=& \int d^{D-1} x\, \Big [ 2\, \Pi^0_{(C)}\, \Pi_{(\bar \beta)} - 2\, \Pi^i_{(\bar C)}\, \Pi^{0i}_{(B)} (b) - \Pi^i_{(C)} \, \partial_i  \beta
 \nonumber\\
 &-& \Pi^{ij}_{(B)} (b) \, \big(\partial_i C_j - \partial_j  C_i \big ) - 2\, \Pi^0_{(\bar C)}\, \Pi_{(\phi)} (b) \Big ] (\vec{x}, t).
\end{eqnarray}
At this stage, we can utilize the well-known relationship between the infinitesimal, continuous 
and nilpotent (anti-)BRST symmetry transformations $s_{(a)b}$ and their 
generators as the Noether conserved (anti-)BRST charges $Q_{(a)b} $ which can be mathematically expressed as
\begin{eqnarray}\label{10}
 s_{(a)b} \, \Phi (\vec{x}, t)\, = \,-\, i\, \Big [\Phi (\vec{x}, t), \; \; Q_{(a)b} \Big ]_{(\pm)},
\end{eqnarray}
where the subscripts $(\pm)$ on the square bracket [cf. the r.h.s. of equation (10)]
denote the (anti)commutator for the generic dynamical field $\Phi$ (of the
coupled Lagrangian densities ${\cal L}_{(\bar B)}$ and ${\cal L}_{(B)}$ [cf. Eq. (1)]) being fermionic/bosonic in nature. To corroborate the 
claim, that has been made in the relationship (10), we work out a couple of examples explicitly  for the 
sake of readers' convenience. For instance, 
we have the following relationships [cf. Eq. (10)] for the components $B_{0i} $ and $B_{ij} $ of the 
{\it bosonic} Abelian 2-form gauge field $B_{\mu\nu}$
\begin{eqnarray}\label{11}
 s_{b} \, B_{0i} (\vec{x}, t)  = -\, i\, \Big [B_{0i} (\vec{x}, t), \; \; Q_{b} \Big ], \qquad
  s_{b} \, B_{ij} (\vec{x}, t)  = -\, i\, \Big [B_{ij} (\vec{x}, t), \; \; Q_{b} \Big ],
\end{eqnarray}
as far as the infinitesimal, continuous and and nilpotent BRST symmetry transformations (2) are concerned. The above relationships can be clearly written as
\begin{eqnarray}\label{12}
 s_{b} \, B_{0i} (\vec{x}, t)  &=& -\, i\, \int d^{D-1} y\,\;\Big [B_{0i} (\vec{x}, t), \;\; -\, 2\, 
\Pi^j_{(\bar C)} (\vec{y}, t)\,  \Pi^{0j}_{(B)} (\vec{y}, t) \Big ] \nonumber\\
&\equiv& 2\, i\, \; \int d^{D-1} y\, \;\Pi^j_{(\bar C)} (\vec{y}, t)\,
\Big [B_{0i} (\vec{x}, t), \; \;\Pi^{0j}_{(B)} (\vec{y}, t) \Big ], \nonumber\\
  s_{b} \, B_{ij} (\vec{x}, t)  &=& -\, i\, \;\int d^{D-1} y\, \Big [B_{ij} (\vec{x}, t), \;\; -\, \Pi^{kl}_{(B)} (\vec{y}, t)\,
\big (\partial_k C_l - \partial_l C_k \big ) (\vec{y}, t) \Big ] \nonumber\\
&\equiv& i\, \int d^{D-1} y\,\; 
\Big [B_{ij} (\vec{x}, t), \; \;\Pi^{kl}_{(B)} (\vec{y}, t) \Big ]\, \big (\partial_k C_l - \partial_l C_k \big ) (\vec{y}, t),
\end{eqnarray}
where we have taken into account {\it only} the appropriate 
and relevant terms of the {\it full} expression for $Q_b$ [cf. Eq. (9)]. At this stage, we use 
the appropriate basic canonical commutators from equation (8) which lead to the following explicit BRST transformations
\begin{eqnarray}\label{13}
s_{b} \, B_{0i} (\vec{x}, t)  &=& -\,  \Pi^i_{(\bar C)} (\vec{x}, t)  \equiv - \big (\partial_0 C_i - \partial_i C_0 \big ) (\vec{x}, t), \nonumber\\
s_{b} \, B_{ij} (\vec{x}, t)  &=& -\, \dfrac{1}{2}\, \big (\delta^k_i \, \delta^l_j - \delta^k_j \, \delta^l_i \big )
 \big (\partial_k C_l - \partial_l C_k \big ) (\vec{x}, t) \equiv - \,\big (\partial_i C_j - \partial_j C_i \big ) (\vec{x}, t)
\end{eqnarray}
which establish that we have already obtained the nilpotent BRST transformation: $s_b B_{\mu\nu} = - (\partial_\mu C_\nu - \partial_\nu C_\mu )$  
because its components are very much present in equation (13).

Now let us focus on the derivation of the BRST symmetry transformation on a {\it fermionic} field: $s_b C_\mu = -\, \partial_\mu \beta$ of our 
BRST-quantized theory. 
Toward this objective in mind, we obtain the following equations from the relationship (10), namely;
\begin{eqnarray}\label{14}
 s_{b} \, C_{0} (\vec{x}, t)  = -\, i\, \Big \{ C_{0} (\vec{x}, t), \; \; Q_{b} \Big \}, \qquad
  s_{b} \, C_{i} (\vec{x}, t)  = -\, i\, \Big  \{C_{i} (\vec{x}, t), \; \; Q_{b} \Big \}.
\end{eqnarray}
The above relationships can be explicitly written as follows:
\begin{eqnarray}\label{15}
 s_{b} \, C_{0} (\vec{x}, t)  &=& -\,2\, i\, \int d^{D-1} y\, \Big \{ C_{0} (\vec{x}, t), \;  \Pi^0_{(C)} (\vec{y}, t)  \Big \} \, 
\Pi_{(\bar \beta)} (\vec{y}, t), \nonumber\\
  s_{b} \, C_{i} (\vec{x}, t)  &=& +\, i\, \int d^{D-1} y\, \Big  \{C_{i} (\vec{x}, t), \;  \Pi^j_{(C)} (\vec{y}, t) \Big \}\, \partial_j \beta (\vec{y}, t).
\end{eqnarray}
Using the appropriate anticommutators from equation (8), we find that
\begin{eqnarray}\label{16}
 s_{b} \, C_{0} (\vec{x}, t)  = 2\, \Pi_{(\bar \beta)} = - \, \dot \beta (\vec{x}, t) \equiv -\, \partial_0 \beta (\vec{x}, t) \qquad
  s_{b} \, C_{i} (\vec{x}, t)  = -\, \partial_i \beta (\vec{x}, t),
\end{eqnarray}
which establish that we have already derived $s_b C_\mu = -\, \partial_\mu \beta$ because 
the BRST symmetry transformations for its components (i.e. $C_0 $ and $C_i $) are present in  equation (16). Thus, it is very clear 
that the relationship quoted in equation (10) is {\it true} for bosonic as well as fermionic fields of our 
D-dimensional BRST-quantized Abelian 2-form gauge theory.

Against the backdrop of the above discussions, we would like to mention that the other off-shell nilpotent BRST symmetry
transformations in equation (2) can {\it also} be obtained by using the relationship (10). These are listed as follows
\begin{eqnarray}\label{17}
 s_{b} \, \beta (\vec{x}, t)  &=& -\, i\, \Big [\beta (\vec{x}, t), \; Q_{b} \Big ] = 0 \qquad
  s_{b} \, \bar \beta (\vec{x}, t)  = -\, i\, \Big [\bar \beta (\vec{x}, t), \;  Q_{b} \Big ] = 2\, \Pi^0_{(C)} \equiv -\, \rho,
  \nonumber\\
 s_{b} \, \phi (\vec{x}, t)  &=& -\, i\, \Big [\phi (\vec{x}, t), \; \; Q_{b} \Big ]  = -\, 2\, \Pi^0_{(\bar C)} \equiv \lambda,
\end{eqnarray}
where we have taken into account the explicit form of the BRST charge from equation (9) and exploited the theoretical strength of the 
{\it basic} canonical (anti)commutators (8). It is straightforward to note that we have: $s_b [\lambda, \; \rho,\; B_\mu ] = 0 $ because
these fermionic as well as bosonic auxiliary fields do {\it not} have their conjugate momenta in $Q_b$ [cf. Eq. (9)]. We would like to
state that {\it exactly} similar kinds of theoretical exercises can be performed with the Noether conserved anti-BRST charge $Q_{ab}$ [cf. Eq. (9)] 
which lead to the derivations of the precise expressions for the anti-BRST symmetry transformations that have been listed in equation (2).
Without going into the details of computations, we list here the key results
\begin{eqnarray}\label{18}
 s_{ab} \, \bar \beta (\vec{x}, t)  &=& -\, i\, \Big [\beta (\vec{x}, t), \; Q_{ab} \Big ] = 0 \qquad s_{ab} \big [\lambda, \; \rho, \; \bar B_\mu \big ] = 0, \nonumber\\
  s_{ab} \, \bar \beta (\vec{x}, t)  &=& -\, i\, \Big [\bar \beta (\vec{x}, t), \;  Q_{ab} \Big ] = 2\, \Pi^0_{(\bar C)} (\vec{x}, t)
\equiv -\, \lambda (\vec{x}, t), \nonumber\\
 s_{ab} \, \phi (\vec{x}, t)  &=& -\, i\, \Big [\phi (\vec{x}, t), \; \; Q_{ab} \Big ]  = -\, 2\, \Pi^0_{(C)} (\vec{x}, t)
\equiv \rho (\vec{x}, t), \nonumber\\
 s_{ab} \, C_{0} (\vec{x}, t)  &=& -\, i\, \Big \{ C_{0} (\vec{x}, t), \; \; Q_{ab} \Big \} = -\,2\,\Pi_{\phi)} (\bar b) (\vec{x}, t)
\equiv \bar B_0 (\vec{x}, t),
 \nonumber\\
  s_{ab} \, C_{i} (\vec{x}, t)  &=& -\, i\, \Big  \{C_{i} (\vec{x}, t), \; \; Q_{ab} \Big \} = 2\, \Pi^{0i}_{(B)} (\bar b) (\vec{x}, t)
\equiv \bar B_i (\vec{x}, t),
\nonumber\\
  s_{ab} \, \bar C_{i} (\vec{x}, t)  &=& -\, i\, \Big  \{\bar C_{i} (\vec{x}, t), \; \; Q_{ab} \Big \} = -\, \partial_i \bar \beta (\vec{x}, t), 
\nonumber\\
 s_{ab} \, \bar C_{0} (\vec{x}, t)  &=& -\, i\, \Big \{ \bar C_{0} (\vec{x}, t), \; \; Q_{ab} \Big \} = 2\, \Pi_{(\beta)} (\vec{x}, t)
 = -\, \partial_0 \beta (\vec{x}, t), \nonumber\\ 
s_{ab} \, B_{0i} (\vec{x}, t)  &=& -\, i\, \Big [B_{0i} (\vec{x}, t), \; \; Q_{ab} \Big ] = \Pi^i_{(C)} (\vec{x}, t) = 
\big (\partial^0 \bar C^i - \partial^i \bar C^0  \big )  \nonumber\\
&\equiv& -\, \big (\partial_0 \bar C_i - \partial_i \bar C_0  \big ) (\vec{x}, t), \nonumber\\
  s_{ab} \, B_{ij} (\vec{x}, t)  &=& -\, i\, \Big [B_{ij} (\vec{x}, t), \; \; Q_{ab} \Big ] = 
-\, \dfrac{1}{2}\, \big (\delta^k_i \, \delta^l_j - \delta^k_j \, \delta^l_i \big )
 \big (\partial_k C_l - \partial_l C_k \big ) (\vec{x}, t) \nonumber\\
&\equiv& - \,\big (\partial_i C_j - \partial_j C_i \big ) (\vec{x}, t),
\end{eqnarray}
which are the off-shell nilpotent 
anti-BRST symmetry transformations (2) for {\it all} the dynamical and auxiliary fields of the 
{\it perfectly} anti-BRST invariant Lagrangian density ${\cal L}_{(\bar B)} $.

We end this subsection with a few concluding remarks. First of all, we note that the Noether conserved (anti-)BRST charges 
[cf. Eqs. (5),(9)] are the generators for the infinitesimal, continuous and off-shell nilpotent (anti-)BRST symmetry transformations (2).
Second, {\it these} charges are {\it not} invariant [cf. Eq. (6)] under the infinitesimal, continuous and off-shell nilpotent 
(anti-)BRST symmetry transformations that have been quoted in equation (2). Finally, a few transformations 
(e.g. $ s_{b}  \bar B_\mu = -\, \partial_\mu \lambda,
\; s_{ab}  B_\mu = -\, \partial_\mu \rho $),
listed in equation (2), have been derived from the requirements of (i) the absolute anticommutativity property
(i.e. $\{s_b, \; s_{ab} \} = 0 $) between the 
off-shell nilpotent BRST and anti-BRST symmetry transformation operators, and (ii) the (anti-)BRST invariance
(i.e. $s_{(a)b} [B_\mu + \bar B_\mu + \partial_\mu \phi ] = 0 $)
of the {\it physical} CF-type restriction on our theory.\\


\section{(Anti-)BRST Invariant Charges: Modified Versions}

Our present section is divided into two subsections. In Subsec. 4.1, we derive the consistently {\it modified} versions
of the conserved (anti-BRST charges $Q_{(A)B}$ [from the Noether conserved (anti-)BRST charges $Q_{(a)b}$] which are {\it invariant} under the
infinitesimal, continuous and off-shell nilpotent (anti-)BRST symmetry transformations (2). Our Subsec. 4.2 deals with the
proof that the charges $Q_{(A)B}$ are indeed {\it invariant} quantities by taking into account (i) the relationship (10) between the 
infinitesimal, continuous  and off-shell nilpotent (anti-)BRST symmetry transformations and their generators as the Noether conserved  
(anti-)BRST charges  $Q_{(a)b}$, and (ii) the equal-time {\it basic} canonical (anti)commutators (8).\\


\subsection{Charges $Q_{(A)B}$ from Charges $Q_{(a)b}$: Explicit Derivations}

We have noted that the Noether conserved (anti-)BRST charges $Q_{(a)b}$ [cf. Eq. (5)]  are {\it not} invariant [cf. Eq. (6)] under the
infinitesimal, continuous and {\it off-shell} nilpotent (anti-) BRST symmetry transformations (2). In our earlier work [32] on the
Noether theorem and nilpotency property of the (anti-)BRST charges, we have proposed a systematic theoretical method to obtain the 
{\it modified} versions of the (anti-)BRST charges $Q_{(A)B} $ (from the Noether conserved (anti-)BRST charges $Q_{(a)b} $)
which are found to be {\it invariant} under the {\it off-shell} nilpotent (anti-)BRST symmetry transformations. In this proposal, we
have used (i) the partial integration along with the Gauss divergence theorem, (ii) the appropriate EL-EoM at suitable places, and
(iii) the application of the off-shell nilpotent (anti-)BRST symmetry transmigrations {\it judiciously}.
We utilize the {\it same} theoretical method [32], first of all, to derive the BRST-invariant (i.e. $s_b \,Q_B = 0$) charge $Q_B$ from the 
Noether conserved BRST charge $Q_b$.
Toward this goal in mind, we focus on the {\it last} term of the expression for the Noether 
BRST charge $Q_b$ [cf. Eq. (5)] which can be re-written as
\begin{eqnarray}\label{19}
-\, \dfrac{1}{2}\, \int d^{D-1} x\; H^{0ij}\, \big (\partial_i C_j - \partial_j C_i \big ) = - \, \int d^{D-1} x\; H^{0ij}\, \partial_i C_j,
\end{eqnarray}
where the antisymmetric (i.e. $H^{0ij} = - \, H^{0ji} $) 
 property of $H^{0ij}$ has been taken into account.
At this juncture, by (i) performing the partial integration, and (ii) throwing away the total {\it space} derivative term due to the validity of Gauss's
divergence theorem, we obtain the {\it modified} mathematical form of equation (19) as follows
\begin{eqnarray}\label{20}
+\, \int d^{D-1} x\; \big (\partial_i H^{0ij} \big ) \,C_j = +\, \int d^{D-1} x\; \big (\partial^0 B^i - \partial^i B^0 \big )\, C_i,
\end{eqnarray}
where we have (i) used the following EL-EoM (that is derived w.r.t. 
the gauge field $B_{\mu\nu}$ from the Lagrangian density ${\cal L}_{(B)} $ [cf. Eq. (1)]), namely;
\begin{eqnarray}\label{21}
\partial_\mu H^{\mu\nu\sigma} + \big (\partial^\nu B^\sigma - \partial^\sigma B^\nu \big ) = 0 \;\;\Longrightarrow \;\;
\partial_i H^{0ij} = +\,  \big (\partial^0 B^j - \partial^j B^0 \big ),
\end{eqnarray}
and (ii) taken into account the choices: $\nu = 0, \; \sigma = j $ in the EL-EoM [cf. Eq. (21)]. The 
final version of the integral (20) will be present in the expression for $Q_B$.
 At this stage,
we apply the BRST transformations (2) on the {\it final} form of the integral (20) which yields:
\begin{eqnarray}\label{22}
s_b \, \Big [\int d^{D-1} x\; \big (\partial^0 B^i - \partial^i B^0 \big )\, C_i\Big ] = -\, 
\int d^{D-1} x\; \big (\partial^0 B^i - \partial^i B^0 \big )\, \partial_i \beta.
\end{eqnarray}
We have to {\it modify} a specific term (existing in the expression for $Q_b$) which can cancel out with (22) when we apply the BRST summery transformation
operator $s_b$ on a part of that term. Such a specific term is the {\it first} term of $Q_b$ [cf. Eq. (5)] which can be {\it modified} as:
\begin{eqnarray}\label{23}
-\, \int d^{D-1} x\; \big (\partial^0 \bar C^i - \partial^i \bar C^0 \big )\, \partial_i \beta  &=& + \, 
\int d^{D-1} x\; \big (\partial^0 \bar C^i - \partial^i \bar C^0 \big )\, \partial_i \beta \nonumber\\
&-&
2\, \int d^{D-1} x\; \big (\partial^0 \bar C^i - \partial^i \bar C^0 \big )\, \partial_i \beta.
\end{eqnarray}
It is crystal clear that when $s_b$ acts on the {\it first} term on r.h.s. of the above equation, it cancels out with the integral in equation (22).
Hence, the {\it first} term, on the r.h.s. of the above equation, will be present in the expression for $Q_B$. Now let us focus on the {\it second}
integral on the r.h.s. of equation (23). This term can be re-expressed as 
\begin{eqnarray}\label{24}
-\, 2\, \int d^{D-1} x\; \big (\partial^0 \bar C^i - \partial^i \bar C^0 \big )\, \partial_i \beta =
+\, 2 \,\int d^{D-1} x\;\, \partial_i \big (\partial^0 \bar C^i - \partial^i \bar C^0 \big )\,  \beta,
\end{eqnarray}
where we have (i) performed  the partial integration, and (ii) used the Gauss divergence theorem to drop the total {\it space} derivative term. Using
the following EL-EoM w.r.t. the field $C_\mu$ (which is derived from the Lagrangian densities ${\cal L}_{(B, \;\bar B)} $ [cf. Eq. (1)]), namely;
\begin{eqnarray}\label{25}
 \partial_\mu \big (\partial^\mu \bar C^\nu - \partial^\nu \bar C^\mu \big ) - \dfrac{1}{2} \, \partial^\nu \rho = 0 
 \;\;\Longrightarrow \; \; \partial_i \big (\partial^0 \bar C^i - \partial^i \bar C^0 \big ) = -\, \dfrac{1}{2}\, \dot \rho,
\end{eqnarray}
we can recast the integral, that is present in equation (24), in the following form
\begin{eqnarray}\label{26}
+ \,2\, \int d^{D-1} x\; \partial_i \big (\partial^0 \bar C^i - \partial^i \bar C^0 \big )\,  \beta =
 -\, \int d^{D-1} x\; \big (\dot \rho \, \beta),
\end{eqnarray}
which interestingly turns out to be  a BRST-invariant (i.e. $s_b [\dot \rho\, \beta] = 0 $) quantity. This integral will be {\it also} present  in the explicit
expression for $Q_B$.  Taking all these inputs, we obtain the BRST-invariant (i.e. $s_b Q_B = 0 $) 
{\it modified} version of the BRST charge $Q_B$ as:
\begin{eqnarray}\label{27}
Q_b \longrightarrow Q_{B} &=& \int d^{D-1} x\,\,
\Big [ \big (\partial^0 B^i - \partial^i B^0 \big )\, C_i +
\big (\partial^0 \bar C^i - \partial^i \bar C^0 \big ) \, \partial_i  \beta \nonumber\\ 
&-& \big (\partial^0  C^i - \partial^i  C^0\big ) \,  B_i 
- \frac{1}{2}\, \rho \, \dot  \beta + \frac{1}{2}\, \lambda \,  B^0 - \dot \rho\, \beta   \Big ]  (\vec{x}, \; t).
\end{eqnarray}
It should be noted that except the {\it two} integrals [which we have consistently modified in equations (19) and (23)], the {\it rest} of the
integrals of $Q_b$ and the integral (26) are BRST-invariant quantities and, therefore, they will be present in the 
precise expression for $Q_B$.
As already pointed out earlier, it is straightforward to check that $s_b Q_B = 0$  where the off-shell nilpotent BRST transformations
are listed in equation (2). This observation is decisively {\it different} from our observation in equation (6) which is connected
with the checking of the BRST invariance of the Noether charge $Q_b$ where we have found that:  $s_b Q_b \neq 0 $.

Now we devote time on a {\it concise} discussion on the derivation of the conserved 
and anti-BRST invariant (i.e. $s_{ab} \, Q_{AB} = 0 $) 
version of the anti-BRST charge $Q_{AB}$ from the Noether anti-BRST charge $Q_{ab}$ [cf. Eq. (5)]. Once again, we perform the 
partial integration and use the Gauss divergence theorem to recast the {\it last} term of $Q_{ab}$ [cf. Eq. (5)] as
\begin{eqnarray}\label{28}
-\, \dfrac{1}{2}\, \int d^{D-1} x\; H^{0ij}\, \big (\partial_i \bar C_j - \partial_j \bar C_i \big ) 
= + \, \int d^{D-1} x\; \big (\partial_i H^{0ij} \big )\; \bar C_j,
\end{eqnarray}
where we have dropped a total {\it space} derivative term.  Using the following EL-EoM w.r.t. the gauge field
from the Lagrangian density ${\cal L}_{(\bar B)} $ [cf. Eq. (1)], namely;
\begin{eqnarray}\label{29}
\partial_\mu H^{\mu\nu\sigma} - \big (\partial^\nu \bar B^\sigma - \partial^\sigma \bar B^\nu \big ) = 0 \;\;\Longrightarrow \;\;
\partial_i H^{0ij} = -\,  \big (\partial^0 \bar B^j - \partial^j \bar B^0 \big ),
\end{eqnarray}
we can recast the integral in (28) as follows:
\begin{eqnarray}\label{30}
= - \, \int d^{D-1} x\; \big (\partial^0 \bar B^i - \partial^i \bar B^0 \big )\;\bar  C_i.
\end{eqnarray}
This integral will be present in the expression for $Q_{AB}$. At this juncture, we apply the anti-BRST 
symmetry transformation operator $s_{ab}$ on the {\it above} integral which leads to:
\begin{eqnarray}\label{31}
s_{ab} \Big [- \, \int d^{D-1} x\; \big (\partial^0 \bar B^i - \partial^i \bar B^0 \big )\;\bar  C_i \Big ] = 
+\, \int d^{D-1} x\; \big (\partial^0 \bar B^i - \partial^i \bar B^0 \big )\;\partial_i \bar \beta.
\end{eqnarray}
As per the proposal in our earlier work [32], we have to {\it modify} a specific term of the Noether anti-BRST charge $Q_{ab}$ [cf. Eq. (5)]
in such a manner that when the anti-BRST symmetry transformation operator $a_{ab}$ acts on a {\it part} of it, the result should cancel
out with the integral in equation (31). Such an appropriate modification is as follows:
\begin{eqnarray}\label{32}
+\, \int d^{D-1} x\; \big (\partial^0 C^i - \partial^i  C^0 \big )\, \partial_i \bar \beta  &=& - \, 
\int d^{D-1} x\; \big (\partial^0  C^i - \partial^i  C^0 \big )\, \partial_i \bar \beta \nonumber\\
&+&
2\, \int d^{D-1} x\; \big (\partial^0  C^i - \partial^i  C^0 \big )\, \partial_i \bar \beta.
\end{eqnarray}
It is straightforward to check that the application of the operator $s_{ab}$ on the {\it first} integral, on the r.h.s. of the above equation,
cancels out with the integral in equation (31). Hence, the integral in equation (30) and the {\it first} integral on the r.h.s. of the 
{\it above} equation
{\it both} will be present in the precise expression for $Q_{AB}$. Let us now concentrate on the {\it second} integral on the r.h.s. of equation (32).
Performing the partial integration and using the theoretical strength of the Gauss divergence theorem, we obtain the following:
\begin{eqnarray}\label{33}
2\, \int d^{D-1} x\; \big (\partial^0  C^i - \partial^i  C^0 \big )\, \partial_i \bar \beta = 
-\, 2\, \int d^{D-1} x\; \partial_i \big (\partial^0  C^i - \partial^i  C^0 \big )\,  \bar \beta.
\end{eqnarray}
At this stage, we exploit the beauty and theoretical strength of the following EL-EoM
\begin{eqnarray}\label{34}
 \partial_\mu \big (\partial^\mu C^\nu - \partial^\nu  C^\mu \big ) + \dfrac{1}{2} \, \partial^\nu \lambda = 0 
 \;\;\Longrightarrow \; \; \partial_i \big (\partial^0  C^i - \partial^i  C^0 \big ) = +\, \dfrac{1}{2}\, \dot \lambda,
\end{eqnarray}
which is  derived from the Lagrangian densities [cf. Eq. (1)] w.r.t. the anti-ghost field $\bar C_\mu$. The substitution of 
equation (34) into equation (33) leads to the following: 
\begin{eqnarray}\label{35} 
-\, 2\, \int d^{D-1} x\; \partial_i \big (\partial^0  C^i - \partial^i  C^0 \big )\,  \bar \beta = 
-\, \int d^{D-1} x\;  \big (\dot \lambda\;   \bar \beta \big ).
\end{eqnarray}
This anti-BRST invariant (i.e. $s_{ab} [\dot \lambda\,   \bar \beta ] = 0 $) integral will {\it also} be present in the precise expression for $Q_{AB}$.
Taking into account {\it all} these inputs {\it together}, we obtain the anti-BRST invariant (i.e. $s_{ab} Q_{AB} = 0 $) version
of the modified anti-BRST  charge $Q_{AB}$ as:
\begin{eqnarray}\label{36}
Q_{ab} \longrightarrow Q_{AB} &=& \int d^{D-1} x\,\,
\Big [\big (\partial^0 \bar  C^i 
- \partial^i  \bar C^0 \big ) \,  \bar B_i -\, \big (\partial^0 \bar B^i - \partial^i \bar B^0 \big )\, \bar C_i \nonumber\\
&-&
\big (\partial^0  C^i - \partial^i C^0 \big ) \, \partial_i  \bar \beta  
+ \frac{1}{2}\, \lambda\, \dot  {\bar \beta} - \frac{1}{2}\, \rho \,  \bar B^0 - \dot \lambda\, \bar \beta   \Big ] (\vec{x}, \; t).
\end{eqnarray}
We re-emphasize that the {\it direct} application of the off-shell nilpotent anti-BRST symmetry transformation operator $s_{ab}$ [cf. Eq. (2)] on
the above expression (for the {\it modified} version of the anti-BRST charge $Q_{AB}$) shows that 
we have obtained: $s_{ab} Q_{AB} = 0 $.

We end this subsection with a couple of concluding remarks. First of all, we note that the applications of (i) the partial integration along with
the Gauss divergence theorem on the term that is associated with the gauge field 
in the expressions for the Noether conserved (anti-)BRST charges, and (ii) the EL-EoM 
w.r.t. the gauge field at suitable places, lead to the derivation
of a {\it specific} term that ought to be present 
in the precise expression for the (anti-)BRST  invariant (i.e. $s_{(a)b} Q_{(A)B} =  0 $)
versions of the (anti-)BRST charges. Furthermore,
we need to apply the {\it combination} of the Gauss divergence theorem and EL-EoM in the ghost-sector 
of our BRST-quantized theory to convert the {\it non-invariant} Noether conserved
(anti-)BRST charges [cf. Eq. (6)] 
into the (anti-)BRST invariant versions of the (anti-)BRST charges $Q_{(A)B}$
which are found to be {\it physical} (in the true sense of the
word) within the framework of  BRST formalism because they remain invariant under the {\it off-shell} nilpotent (anti-)BRST symmetry transformations.
Second, as  far as the physicality criteria (within the framework of BRST formalism) are concerned, we observe that {\it only} the
(anti-)BRST invariant versions of the (anti-)BRST charges lead to the results that are consistent with the Dirac quantization conditions for
the gauge theories that are endowed with the  first-class constraints [25-31]. We would like to add  here that 
the Noether conserved (anti-)BRST charges $Q_{(a)b}$ 
lead to  {\it absurd} results if they are used in the physicality criteria within the framework of BRST formalism (cf. Sec. 5 for more
discussions).\\


\subsection{(Anti-)BRST Invariance of $Q_{(A)B}$: Basic Brackets}

In this subsection, we prove the (anti-)BRST invariance (i.e. $s_{(a)b} Q_{(A)B} =  0 $) of the {\it modified} versions of the 
(ant-)BRST charges $Q_{(A)B}$ by taking into considerations the theoretical strength of the basic canonical (anti)commutators 
that have been quoted in our equation (8). Toward this goal in our mind, first of all, we express the {\it modified} version of the
ant-BRST charge $Q_{AB}$ in terms of the canonical conjugate momenta that have been defined in our equation (7) for
the present BRST-quantized theory. It turns out that we have the following form of the {\it modified} version of 
the anti-BRST charge $Q_{AB}$, namely;
\begin{eqnarray}\label{37}
 Q_{AB} &=& \int d^{D-1} x\, \Big [ 2\, \Pi^0_{(\bar C)}\, \Pi_{(\beta)} 
+ 2\, \Pi^i_{(C)}\, \Pi^{0i}_{(B)} (\bar b) + \Pi^i_{(\bar C)} \, \partial_i \bar \beta
 \nonumber\\
 &-&  2\, \Pi^0_{(C)}\, \Pi_{(\phi)} (\bar b) - \dot \lambda\, \bar \beta 
- \big (\partial^0 \bar B^i -  \partial^i \bar B^0 \big )\, \bar C_i \Big ] (\vec{x}, \; t), 
\end{eqnarray}
which is the analogue of our earlier expression for the {\it Noether} conserved anti-BRST charge $Q_{ab}$ 
in terms of the canonical conjugate momenta [cf. Eq. (9)]. At this stage, we exploit the beauty and theoretical strength of
the relationship in equation (10) to prove that
\begin{eqnarray}\label{38}
 s_{ab} \, Q_{AB}   = \,-\, i\, \Big \{ Q_{AB}, \; \; Q_{ab} \Big \}  = 0,
\end{eqnarray}
where we have chosen, in the {\it general} kind of 
relationship [cf. Eq. (10)],
 the generic field $\Phi$  to be (i.e. $\Phi = Q_{AB} $)
the {\it modified} version of the anti-BRST charge $Q_{AB}$  itself.

We are in the position now to prove that $  s_{ab} \, Q_{AB} = 0$ by demonstrating that the r.h.s. (i.e. $\{ Q_{AB}, \; \; Q_{ab} \}  = 0 $) 
of the above equation (38) is indeed {\it zero} if we use (i) the appropriate {\it basic} canonical (anti)commutators from equation (8)
that are valid for our BRST-quantized theory, and (ii) the suitable EL-EoM that is derived from the Lagrangian density ${\cal L}_{(\bar B)} $ [cf. Eq. (1)].
In its full blaze of glory, the anticommutator $\{ Q_{AB}, \; \; Q_{ab} \} $ looks as  follows
\begin{eqnarray}\label{39}
 &&\Big \{ Q_{AB}, \; \; Q_{ab} \Big \}  = \int \int d^{D-1} x\, d^{D-1} y\,\Big \{ \Big ( 2\, \Pi^0_{(\bar C)}\, \Pi_{(\beta)} 
+ 2\, \Pi^i_{(C)}\, \Pi^{0i}_{(B)} (\bar b) + \Pi^i_{(\bar C)} \, \partial_i \bar \beta
 \nonumber\\
 &-&  2\, \Pi^0_{(C)}\, \Pi_{(\phi)} (\bar b) - \dot \lambda\, \bar \beta 
- \big (\partial^0 \bar B^i -  \partial^i \bar B^0 \big )\, \bar C_i \Big ) (\vec{x},\; t), \; \; 
\Big ( 2\, \Pi^0_{(\bar C)}\, \Pi_{(\beta)} + 2\, \Pi^j_{(C)}\, \Pi^{0j}_{(B)} (\bar b)  \nonumber\\
&-& \Pi^j_{(\bar C)} \, \partial_j \bar \beta
- \Pi^{kl}_{(B)} (\bar b) \, \big(\partial_k \bar C_l - \partial_l \bar C_k \big ) - 2\, \Pi^0_{(C)}\, \Pi_{(\phi)} (\bar b) \Big ) (\vec{y},\; t)
 \Big \},
\end{eqnarray}  
where {\it all} the symbols have already been explained in our earlier discussions.
A close and careful  look at the above equation, in view of the 
rules of the {\it basic} canonical brackets in equation (8),
 shows that the {\it non-zero} exiting integrals are as follows:
\begin{eqnarray}\label{40}
 &&\Big \{ Q_{AB}, \; \; Q_{ab} \Big \}  = -\, \int \int d^{D-1} x\, d^{D-1} y\,\Big \{ \Big ( \Pi^i_{(\bar C)} \, \partial_i \bar \beta \Big ) (\vec{x}, t), \; 
\Big (\Pi^{kl}_{(B)}  \, \big [\partial_k \bar C_l - \partial_l \bar C_k \big ] \Big ) (\vec{y}, t)
 \Big \} \nonumber\\
&& +  \int \int d^{D-1} x\, d^{D-1} y\, \Big \{ \Big (
 \big [\partial^0 \bar B^i -  \partial^i \bar B^0 \big ]\, \bar C_i \Big ) (\vec{x},\; t), \; \; 
\Big ( \Pi^j_{(\bar C)} \, \partial_j \bar \beta \Big ) (\vec{y},\; t) \Big \}.
\end{eqnarray}
Using the standard rules of the (anti)commutators with the composite operators and/or independent operators
(cf. Appendix E for details), we end-up with the following
\begin{eqnarray}\label{41}
 &&\Big \{ Q_{AB}, \; \; Q_{ab} \Big \}  = -\, \int \int d^{D-1} x\, d^{D-1} y\; H^{0kl} (\vec{y}, t) \,
\Big \{ \Pi^i_{(\bar C)} \,  (\vec{x}, t), \; (\partial_k \bar C_l) (\vec{y}, t)
 \Big \} \,\partial_i \bar \beta (\vec{x}, t) \nonumber\\
&& +  \int \int d^{D-1} x\, d^{D-1} y\; 
\big [\partial^0 \bar B^i -  \partial^i \bar B^0 \big ] (\vec{x}, t)\, \Big \{ \bar C_i  (\vec{x},\; t), \;  
\Pi^j_{(\bar C)}  (\vec{y},\; t) \Big \}\, \partial_j \bar \beta (\vec{y},\; t),
\end{eqnarray}
where we have (i) used the expression for the
canonical momenta: $\Pi^{ij}_{(B)} = \frac{1}{2}\, H^{0ij} $ [cf. Eq. (7)],
and (ii) taken care of the totally antisymmetric nature of $H^{0ij}$. Exploiting the appropriate form of the
{\it basic} canonical anticommutators [cf. Eq. (8)],
we obtain the {\it final} expression for the above anticommutator (i.e. $\{ Q_{AB}, \; Q_{ab} \} $) as 
\begin{eqnarray}\label{42}
 \Big \{ Q_{AB}, \;  Q_{ab} \Big \}  &=& +\, i\, \int  d^{D-1} x\, \; \Big ([\partial_k H^{0kl}]  \; \partial_l \bar \beta \Big ) \,(\vec{x},\; t)
 \nonumber\\
 &+& i\,  \int d^{D-1} x\, 
\Big (\big [\partial^0 \bar B^i -  \partial^i \bar B^0 \big ] \, \partial_i \bar \beta \Big ) (\vec{x},\; t),
\end{eqnarray}
where we have (i) performed the partial integration over the volume integral w.r.t. the $y$ variable, (ii) exploited the beauty and theoretical strength of the
Gauss divergence theorem, (iii) used the property of the Dirac $\delta$-function, and (iv) taken
care of the derivative on the Dirac $\delta$-function while carrying out the partial integration. Taking the {\it sum} of the integrals in
the above equation {\it together}, we obtain the following equation
\begin{eqnarray}\label{43}
 \Big \{ Q_{AB}, \;  Q_{ab} \Big \}  &=& +\, i\, \int  d^{D-1} x\, \; \Big (\partial_i H^{0ij} 
 + \big [\partial^0 \bar B^j -  \partial^j \bar B^0 \big ] \Big )\, \partial_j \bar \beta  = 0,
\end{eqnarray}
where we have used the EL-EoM [cf. Eq. (29)] 
w.r.t. the antisymmetric
gauge field $B_{\mu\nu}$ from the Lagrangian density ${\cal L}_{(\bar B)} $ [cf. Eq. (1)]. Thus,
we conclude that the anticommutator: $\{ Q_{AB}, \; Q_{ab} \}$  turns out to be {\it zero} provided we make use of
(i) the partial integration along with the Gauss divergence theorem, and (ii) the appropriate EL-EoM at suitable places.

Against the backdrop of the above discussions related to the anti-BRST charge $Q_{AB}$, we exploit exactly the same 
theoretical technique to the prove the BRST invariance 
(i.e. $s_b Q_B = -\, i\, \{ Q_B,\; Q_b \} = 0  $) of the {\it modified} version of the BRST charge $Q_B$ where we take into account
the equal-time {\it basic} (anti)commutators [cf. Eq. (8)] of our BRST quantized theory in the explicit proof of $\{ Q_B,\; Q_b \} = 0 $.
We would like to be very brief in our forthcoming discussions (in view of our detailed discussions for
the anti-BRST invariance of the {\it modified} version of the anti-BRST charge $Q_{AB}$).
Written in its full blaze of glory, the explicit form of the anticommutator $\{ Q_B,\; Q_b \}$ is as follows
\begin{eqnarray}\label{44}
 &&\Big \{ Q_{B}, \; \; Q_{b} \Big \}  = \int \int d^{D-1} x\, d^{D-1} y\,\Big \{ \Big (\big [\partial^0 B^i - \partial^i B^0 \big ]\, C_i +
\Pi^i_{(C)}\, \partial_i  \beta - 2\, \Pi^i_{(\bar C)}  \,  \Pi^{0i}_{(B)} (b) \nonumber\\
&-& \dot \rho\, \beta
- 2\, \Pi^0_{(C)} \, \Pi_{(\bar \beta)}  + 2\, \Pi^0_{(\bar C)} \, \Pi_{(\phi)} (b \Big ) (\vec{x},\; t), \; \; 
\Big ( 2\, \Pi^0_{(C)}\, \Pi_{(\bar \beta)} - 2\, \Pi^j_{(\bar C)}\, \Pi^{0j}_{(B)} (b) - \Pi^j_{(C)} \, \partial_j  \beta
 \nonumber\\
 &-&  
 \Pi^{kl}_{(B)} (b) \, \big [\partial_k C_l - \partial_l  C_k \big ] - 2\, \Pi^0_{(\bar C)}\, \Pi_{(\phi)} (b) \Big ) (\vec{y},\; t)
 \Big \},
\end{eqnarray} 
where we have expressed the {\it modified} version of the BRST charge $Q_B$ [cf. Eq. (27)] in terms of the canonical conjugate
 momenta [cf. Eq. (7)], namely; 
\begin{eqnarray}\label{45}
Q_{B} &=& \int d^{D-1} x\,\,
\Big [ \big (\partial^0 B^i - \partial^i B^0 \big )\, C_i +
\Pi^i_{(C)}\, \partial_i  \beta - 2\, \Pi^i_{(\bar C)}  \,  \Pi^{0i}_{(B)} (b) - \dot \rho\, \beta \nonumber\\
&-& 2\, \Pi^0_{(C)} \, \Pi_{(\bar \beta)}  + 2\, \Pi^0_{(\bar C)} \, \Pi_{(\phi)} (b)   \Big ],
\end{eqnarray}
that emerge out from the Lagrangian density ${\cal L}_{(B)}$ [cf. Eq. (1)]. All the symbols (that are present in the above equation) have
already been explained earlier in our discussion after equation (7).
As far as the expression for $Q_b$ is concerned, we have
taken the help of equation (9). A close and careful look at the above equation [cf. Eq. (44)] 
shows that the following non-zero existing anticommutators  
are relevant for our purpose, namely;
\begin{eqnarray}\label{46}
 &&\Big \{ Q_{B}, \; \; Q_{b} \Big \}  = -\, \int \int d^{D-1} x\, d^{D-1} y\; H^{0kl} (\vec{y}, t) \,
\Big \{ \Pi^i_{(C)} \,  (\vec{x}, t), \; (\partial_k  C_l) (\vec{y}, t)
 \Big \} \,\partial_i \ \beta (\vec{x}, t) \nonumber\\
&& -  \int \int d^{D-1} x\, d^{D-1} y\; 
\big [\partial^0  B^i -  \partial^i  B^0 \big ] (\vec{x}, t)\, \Big \{  C_i  (\vec{x},\; t), \;  
\Pi^j_{(C)}  (\vec{y},\; t) \Big \}\, \partial_j  \beta (\vec{y},\; t).
\end{eqnarray}
Using the equal-time {\it basic} canonical anticommutator [cf. Eq. (8)] of our present BRST-quantized theory, we obtain, ultimately, the following
result in the integral form, namely;
\begin{eqnarray}\label{47}
s_b Q_B = -\, i\,  \Big \{ Q_{B}, \;  Q_{b} \Big \}  =  \int  d^{D-1} x\, \; \Big (\partial_i H^{0ij} 
 - \big [\partial^0  B^j -  \partial^j  B^0 \big ] \Big )\, \partial_j  \beta  = 0.
\end{eqnarray}
In the {\it above}
 claim, we have taken into account the EL-EoM [cf. Eq. (21)] that is derived from the Lagrangian density ${\cal L}_{(B)}$ [cf. Eq. (1)]
w.r.t. the antisymmetric tensor gauge field $B_{\mu\nu}$. To sum-up, we have proven the BRST-invariance (i.e. $s_b Q_B = -\, i\, \{ Q_B,\; Q_b \} = 0  $)
of the {\it modified} version of the BRST charge $Q_B$ by (i) the direct application of the infinitesimal, continuous and off-shell nilpotent
BRST symmetry transformation operator $s_b$ on the explicit expression for $Q_B$ [cf. Eq. (27)], and (ii)
explicitly demonstrating that the anticommutator $ \{ Q_B,\; Q_b \} $ is {\it zero}
provided we use $(a)$ the partial integration along with the Gauss divergence theorem, $(b)$ the appropriate EL-EoM at suitable places, and
$(c)$ the equal-time {\it basic} canonical anticommutators [cf. Eq. (8)] of our BRST-quantized theory.\\


\section{Physicality Criteria: $Q_{(a)b}$ Versus $Q_{(A)B}$}

Our present section is divided into two subsections. In Subsec. 5.1, we discuss the physicality criteria 
(i.e. $Q_{(a)b} \; |phys> = 0 $) w.r.t. the Noether
conserved  (anti-)BRST charges $Q_{(a)b}$ where the {\it true} physical state of our theory is denoted by the symbol: $|phys> $.
Our Subsec. 5.2 is  devoted to the discussion on the  physicality criteria 
(i.e. $Q_{(A)B} \; |phys> = 0 $) w.r.t. the
conserved and (anti-)BRST invariant versions of  the (anti-)BRST charges $Q_{(A)B}$.\\


\subsection{Physicality Criteria: Noether (Anti-)BRST Charges}

The central objective of our present subsection is to show that when we take the Noether conserved (anti-)BRST charges
$Q_{(a)b}$ in the physicality criteria (i.e. $Q_{(a)b} \; |phys> = 0 $), we obtain {\it absurd} results in the
sense that the conditions on the physical states do {\it not} agree with the Dirac quantization conditions  that
ought to be imposed on the
physical states  of the quantized versions of the 
gauge theories (that are endowed with the first-class constraints). To be precise, we observe that the
physicality criteria (i.e. $Q_{(a)b} \; |phys> = 0 $) lead to the annihilation of the physical state  $|phys> $
by the components (i.e. $\Pi^{0i}_{(B)}, \; \Pi^{ij}_{(B)} $)
of the canonical conjugate momenta $\Pi^{\mu\nu}_{(B)}$ [cf. Eq. (7)] that are derived from the {\it coupled}
Lagrangian densities ${\cal L}_{(B)}$ and ${\cal L}_{(B)}$ [cf. Eq. (1)] w.r.t. the antisymmetric tensor gauge field $B_{\mu\nu}$.
This observation is in sheer contrast to the Dirac quantization conditions on the physical states $|phys> $
for the quantized versions of the gauge theories
where the operator forms of the first-class constraints  are {\it required} to annihilate the physical states [26].

To corroborate the above claims, we would like to begin with a very careful look at the {\it coupled} 
Lagrangian densities ${\cal L}_{(B)}$  and ${\cal L}_{(\bar B)}$ [cf. Eq. (1)] where we observe that, right from the beginning, the
physical fields (carrying the ghost number equal to {\it zero}) and the FP-ghost fields (carrying the {\it non-zero} ghost numbers)
are {\it decoupled}. As a consequence, a specific quantum state in the total quantum Hilbert space of states is a direct product 
(see, e.g. [33-36] for details) of the physical states (i.e. $|phys> $) and the ghost states (i.e.  $|ghost> $). When the 
operator forms of the Noether conserved (anti-)BRST charges operate on the quantum states of the Hilbert space, the
physical fields (with zero ghost numbers) act on the physical states (i.e. $|phys> $) and the (anti-)ghost fields
(with non-zero ghost numbers) operate on the ghost states ((i.e.  $|ghost> $). The {\it latter} operation always
yields {\it non-zero} result. The physicality criteria (i.e. subsidiary conditions), within the framework of BRST formalism,
state that the {\it true} physical states (existing in the total quantum Hilbert space of states) 
are {\it those} that are annihilated (i.e. $Q_{(a)b} \; |phys> = 0 $)
by the operator forms of the (anti-)BRST charges. In other words, these criteria require that {\it all} the physical fields
(with ghost number equal to zero) must annihilate the physical states provided they are associated with the {\it basic}
(anti-)ghost fields of our theory.

In view of the above discussions, first of all, let us focus on the explicit expression for the Noether conserved 
BRST charge [cf. Eq. (5)] and perform the thread-bare analysis of  the individual terms that are present in $Q_b$.
We find that (i) every individual term carnies an {\it effective} ghost number equal to $+\,1$, and (ii) there
are {\it two} types of terms that are present in the explicit expression for $Q_b$. 
As far as the {\it latter} point is concerned, we note that the {\it first} type
of terms are those that are made up of {\it only} the (anti-)ghost fields [e.g. $+\, \frac{1}{2}\, \rho\, \dot \beta, \;
- (\partial^0 \bar C^i - \partial^i \bar C^0)\, \partial_i \beta $] with the {\it effective} ghost number equal to $+\,1$. The 
{\it second} type of terms [e.g. $-\, \frac{1}{2}\, \lambda\, B^0, \; - (\partial^0  C^i - \partial^i  C^0)\, B_i, \;
- \, \frac{1}{2}\, H^{0ij}\, (\partial_i C_j - \partial_j C_i) $], on the other hand,  are those that contain fields 
(with ghost number equal to zero) and the ghost fields with ghost number equal to  $+\,1$. As argued earlier, the 
{\it first} kinds of terms will act only on the ghost states (i.e.  $|ghost> $)
and produce the non-zero results. These terms will {\it not} contribute anything to the physicality criterion w.r.t.
the BRST charge $Q_b$. As far as the {\it second} type of terms are concerned, the ghost fields (with the ghost number equal to $+\,1$) would act 
{\it only } on the ghost states (i.e.  $|ghost> $)
and produce the non-zero results. Thus, to satisfy the subsidiary condition (i.e. physicality criterion $Q_{b} \; |phys> = 0$
w.r.t. the Noether BRST charge $Q_b$),
{\it only} the fields (with ghost number equal to zero) would act on the  physical states (i.e. $|phys> $) and would lead to 
the specific set of conditions on them.

Against the backdrop of the above discussions, let us actually perform the operation: $Q_{b} \; |phys> = 0$ 
(w.r.t. the Noether BRST charge $Q_b$) explicitly and observe the 
consequences that ensue from the above criterion. To be precise, this operation leads to 
\begin{eqnarray}\label{48}
B_i \; |phys> &=& 0 \;\;\;\Longrightarrow \;\;\; - \,2\,\Pi^{0i}_{(B)}\; |phys> = 0,   \nonumber\\
\dfrac{1}{2}\, H^{0ij}\, |phys> &=& 0 \;\;\;\Longrightarrow \;\;\; \Pi^{ij}_{(B)}\; |phys> = 0,
\end{eqnarray}
because of the fact that the fields $B_i$ and $\frac{1}{2}\, H^{0ij} $ are associated with the specific combination of derivatives on
the components (i.e. $C_0, \; C_i $) of the {\it basic} ghost field $C_\mu$. However, we note that the conditions obtained in the
above equation (48) are {\it not} consistent with the Dirac quantization conditions on the quantized versions of the gauge theories
where the operator forms of the first-class constraints are {\it required} to annihilate the {\it true} physical states (i.e. $|phys> $).
To be precise, the results in equation (48) are {\it absurd}. Whereas the top entry [cf. Eq. (48)] implies that the operator
form of the PC [cf. Eq. $(C.2)$ below] annihilates the physical states, the bottom entry implies that the operator form of the {\it non-zero} 
canonical conjugate momenta $\Pi^{ij}_{(B)} $ [cf. Eq. (7)] should {\it also} annihilate the physical states. This
condition  is absolutely wrong
because $ \Pi^{ij}_{(B)} = \frac{1}{2}\, H^{0ij}$ is {\it not} a constraint on our theory. Exactly the {\it same} conditions are obtained
when we discuss the physicality criterion (i.e. $Q_{ab} \; |phys> = 0$) w.r.t. the 
Noether conserved anti-BRST charge $Q_{ab}$ [cf. Eq. (5)]. Thus, we conclude that the Noether conserved (anti-)BRST charges $Q_{(a)b}$ 
are {\it not} suitable for the physicality criteria. This happens because of the fact that {\it these} conserved charges
are {\it not} invariant [cf. Eq. (6)] under the {\it off-shell} nilpotent (anti-)BRST symmetry transformations (2). Hence, these 
Noether conserved (anti-)BRST charges
are {\it not} physical quantities (in the true sense of the word) within the framework of BRST formalism.

We end this subsection with a couple of crucial remarks. First of all, we note that 
 the term ($-\, \frac{1}{2}\, \lambda \, B^0 $), present in the expression for $Q_b$
[cf. Eq. (5)], does {\it not} contribute anything to the physicality criterion  (i.e. $Q_{b} \; |phys> = 0$) because the $B^0$ field
(carrying the ghost number equal to zero) is associated with an auxiliary fermionic field [i.e. $\lambda = +2\, (\partial \cdot C) $]
which is {\it not}  the basic ghost field of our BRST-quantized theory. Furthermore, we note that the canonical conjugate momentum 
$\Pi_{(\phi)} (b) = - \, \frac{1}{2}\, B^0$
w.r.t. the scalar field $\phi$ [cf. Eq. (7)] is {\it not}  a constraint on our BRST-quantized theory. This is why we do {\it not} have 
condition like: $B^0 \, |phys> = 0 $ on our D-dimensional BRST-quantized version of the Abelian 2-form gauge theory. Second, all
the arguments and discussions connected with the Noether conserved BRST charge $Q_b$ turn out to be the {\it same} in the context of the 
physicality criterion (i.e. $Q_{ab} \; |phys> = 0$)  w.r.t. the Noether conserved anti-BRST charge $Q_{ab}$. In other words, we do {\it not}
obtain a condition like: $\bar B^0\, |phys> = 0 $ because the field $\bar B^0$ is associated with an auxiliary fermionic
field [i.e. $\rho = -\, 2\, (\partial \cdot \bar C) $]  in the explicit expression for the Noether anti-BRST charge $Q_{ab}$ [cf. Eq. (5)]
which is {\it not} the basic anti-ghost field of our theory.\\


\subsection{Physicality Criteria: Modified (Anti-)BRST Charges}

In the previous subsection, we have discussed thoroughly the physicality criteria (i.e. $Q_{(a)b} \; |phys> = 0$)  w.r.t. the 
Noether conserved (anti-)BRST charges $Q_{(a)b}$ which are {\it not} invariant under the infinitesimal, continuous and {\it off-shell}
nilpotent (anti-)BRST symmetry transformation operators $s_{(a)b}$ [cf. Eq. (6)]. In view of this detailed discussions, we shall 
very concisely discuss the physicality criteria (i.e. $Q_{(A)B} \; |phys> = 0$) w.r.t. the {\it modified}  versions of the
(anti-)BRST charges $Q_{(A)B} $ which are (i) the conserved quantities because they have been derived from the 
Noether conserved (anti-)BRST charges $Q_{(a)b}$ by exploiting the theoretical strength of the EL-EoM along
with the Gauss divergence theorem at appropriate places, and (ii) the (anti-)BRST invariant (i.e. $s_b Q_B = 0, \; s_{ab} Q_{AB} = 0 $) 
quantities  under the 
{\it off-shell} nilpotent (anti-)BRST symmetry transformations (2). As a consequence, these {\it modified} versions of the 
(anti-)BRST charges  $Q_{(A)B} $ are the {\it physical} quantities within the framework of BRST formalism.

Against the backdrop of the above paragraph, let us discuss the physicality criterion (i.e. $Q_{B} \; |phys> = 0$ w.r.t. the
{\it modified} version of the BRST charge $Q_B$ [cf. Eq. (27)]. We invoke the same logical arguments that  have been
thoroughly discussed in the previous Subsec. 5.1. With that theoretical background, we note that the physicality criterion
$Q_B\, |phys> = 0 $ (w.r.t. the {\it modified} BRST invariant: $s_b Q_B = 0 $ BRST charge $Q_B$), we have the following 
conditions on the physical states (i.e. $|phys> $), namely;
\begin{eqnarray}\label{49}
 B_i \; |phys> = 0 \;\;\;&\Longrightarrow& \;\;\; - \,2\,\Pi^{0i}_{(B)}\; |phys> = 0,   \nonumber\\
\big (\partial^0 B^i - \partial^i B^0 \big )\, |phys> = 0 \;\;\;&\Longrightarrow& \;\;\;
(\partial_j \, H^{0ji}) \, |phys> = 0 \nonumber\\
& \equiv & 2\, (\partial_j \Pi^{ji}_{(B)}) \; |phys> = 0,
\end{eqnarray}
where we have used (i) the expression for the conjugate momenta: $\Pi^{ij}_{(B)} = \frac{1}{2}\, H^{0ij} $, and
(ii) the EL-EoM: $  \partial_i H^{0ij} = +\,  \big (\partial^0 B^j - \partial^j B^0 \big ) $ [cf. Eq. (21)]. 
A close and careful look at the above equation (49) shows that the requirement: $Q_{B} \; |phys> = 0$ (w.r.t.
the conserved and BRST-invariant version of the BRST charge $Q_B$) leads to the annihilation of the physical states (i.e. $|phys> $)
by the operator forms of the first-class constraints (which have been explicitly discussed in our Appendix C at the {\it classical} level).
Thus, the observations in equation (49) are consistent with the Dirac quantization conditions on the gauge theories 
where the essential requirement is the annihilation of the physical states by the operator forms of the first-class constraints
(see, e.g. [26] for more discussions).

We wrap-up this short but very important subsection with a couple of  remarks. First, the specific term (i.e. $-\, \frac{1}{2}\, \lambda\, B^0 $),
present in the expression for the {\it modified} BRST charge $Q_B$)
does {\it not} contribute anything to the physicality criterion; $Q_{B} \; |phys> = 0$ because the physical field $B^0$ 
(with the ghost number equal to zero) is associated
with the fermionic auxiliary field [i.e. $\lambda = +\, 2\, (\partial \cdot C) $] which is {\it not} the basic ghost field
of our BRST-quantized theory. Second, we obtain exactly the same type of conditions as (49) from the physicality criterion:
 $Q_{AB} \; |phys> = 0$ w.r.t. the anti-BRST invariant (i.e. $s_{ab} Q_{AB} = 0 $) and  {\it modified} version
 of the conserved anti-BRST charge $Q_{AB}$ [cf. Eq. (36)]. In other words, the subsidiary condition (i.e. physicality criterion) w.r.t. the
 {\it modified} anti-BRST charge $Q_{AB}$ also leads to the annihilation of the {\it true} physical states (i.e. $|phys> $)
by the operator forms of the first-class constraints (cf. Appendix C for details)
which are found to be consistent with the Dirac  quantization conditions
on the quantized versions of the gauge theories.\\


\section{ Nilpotency Property of $Q_{(a)b}$: Nilpotent Symmetry Considerations and Basic Brackets}

This section is divided into two parts. In Subsec. 6.1, we concentrate on the standard relationship between the infinitesimal 
and continuous symmetry transformations and their generators as the Noether conserved charges. We prove the nilpotency 
(i.e. $Q_{(a)b}^2 = 0 $) as well as
the (anti-)BRST invariance (i.e. $s_{ab} Q_{ab} = 0, \;s_{b} Q_{b} = 0 $)
of the Noether conserved (anti-) BRST charges $Q_{(a)b}$ by exploiting the theoretical strength
of (i) the appropriate EL-EoM 
at suitable places, and (ii) the Gauss divergence theorem. Our Subsec. 6.2 is devoted to the explicit computations
of the anticommutators $\{Q_b, \; Q_b \} = \frac{1}{2}\, Q_b^2  $ and $\{Q_{ab}, \; Q_{ab} \} = \frac{1}{2}\, Q_{ab}^2  $ 
by exploiting the beauty of the equal-time basic
canonical (ant)commutators that have been listed in our equation (8) [thereby  proving the nilpotency 
(i.e. $Q_{(a)b}^2 = 0 $) property].\\


\subsection{ Nilpotency Property: Symmetry Considerations}

We have already derived the Noether conserved (anti-)BRST charges [cf. Eq. (5)] and shown that these charges are
{\it not} invariant [cf. Eq. (6)] under the infinitesimal, continuous and {\it off-shell} nilpotent (anti-)BRST symmetry
transformations (2). The key objective of our present subsection is (i) to comment on our observations in equation (6)
in view of the relationship between the infinitesimal and continuous symmetry transformations and their generators
as the Noether conserved charges, and (ii) to show that the use of appropriate EL-EoM along with the 
theoretical strength of Gauss divergence theorem, lead to the proof that the Noether conserved (anti-)BRST charges
are $(a)$ the (anti-)BRST invariant (i.e. $s_{ab} Q_{ab} = 0, \; s_b Q_b = 0$) quantities, and $(b)$ the nilpotent (i.e. $Q_{(a)b}^2 = 0$)  
of order two. In this context, first of all, we recast our equation (6) in the following forms: 
\begin{eqnarray}\label{50}
s_{ab} Q_{ab} &=& -\,i\, \Big \{Q_{ab}, \; Q_{ab} \Big \} \equiv
\int d^{D-1} x\, \Big [\big (\partial^0 \bar B^i - \partial^i \bar B^0 \big ) \, \partial_i \bar \beta \Big ] \neq 0, \nonumber\\
s_b Q_{b} &=& -\,i\, \Big \{Q_{b}, \; Q_{b} \Big \} 
\equiv - \int d^{D-1} x\, \Big [\big (\partial^0  B^i - \partial^i  B^0 \big ) \, \partial_i  \beta \Big ] \neq 0.
\end{eqnarray}
It should be pointed out, at this stage,  that the integrals on the extreme r.h.s. of the above equation are nothing but the explicit
computations of the l.h.s. (i.e. $s_{ab} Q_{ab}, \; s_b Q_b $)
of the above equation by {\it direct} applications of the infinitesimal, continuous and 
{\it off-shell} nilpotent (i.e. $s_{(a)b}^2 = 0 $) versions of the (anti-)BRST 
transformations (2) on the precise expressions for the Noether conserved (anti-)BRST charges $Q_{(a)b}$ [cf. Eq. (5)].

At this stage, we invoke the EL-EoM (21) and (29) that have been derived from the {\it coupled} 
Lagrangian densities ${\cal L}_{(B)}$  and ${\cal L}_{(\bar B)}$ [cf. Eq. (1)] w.r.t. the antisymmetric tensor gauge field $B_{\mu\nu}$,
respectively. We note that the r.h.s. of the above equation (50) become
\begin{eqnarray}\label{51}
s_{ab} Q_{ab} &=& -\,i\, \Big \{Q_{ab}, \; Q_{ab} \Big \} \equiv
-\, \int d^{D-1} x\,\; \Big [\big (\partial_i H^{0ij} \big ) \, \partial_j \bar \beta \Big ]  \nonumber\\
s_b Q_{b} &=& -\,i\, \Big \{Q_{b}, \; Q_{b} \Big \} 
\equiv - \int d^{D-1} x\,\; \Big [\big ( \partial_i H^{0ij}\big ) \, \partial_j  \beta \Big ],
\end{eqnarray}
if we substitute the EL-EoM (29) and (21), respectively. In other words, we have obtained the integrand on 
the r.h.s. of the equation (50) as the total
{\it space} derivatives as follows
\begin{eqnarray}\label{52}
s_{ab} Q_{ab} &=& 
-\, \int d^{D-1} x\, \; \partial_i \Big [ H^{0ij}  \, \partial_j \bar \beta \Big ]  \; \;\;\Longrightarrow \; \; \; 0,\nonumber\\
s_b Q_{b} &=& 
 - \int d^{D-1} x\, \;\partial_i  \Big [  H^{0ij} \, \partial_j  \beta \Big ] \; \; \;\Longrightarrow \; \;\; 0,
\end{eqnarray}
where we have exploited the theoretical strength of the Gauss divergence theorem due to which the volume integral gets
converted int the surface integral and all the physical fields vanish off  when $x \to \pm\, \infty $. The above observations
establish that the Noether conserved (anti-)BRST charges $Q_{(a)b}$ are (i) the (anti-)BRST invariant (i.e. $s_{ab} Q_{ab} = 0, \;
s_b Q_b = 0 $) quantities, and (ii) the nilpotent (i.e. $ Q_{ab}^2 = \frac{1}{2}\, \{Q_{ab}, \; Q_{ab} \} = 0,\;
Q_{b}^2 = \frac{1}{2}\, \{Q_{b}, \; Q_{b} \}  = 0$) quantities, too [cf. Eqs. (50, (51)], provided we exploit the
theoretical strength of (i) the relationships between the infinitesimal, continuous and off-shell nilpotent (anti-)BRST
symmetry transformations and their generators as the Noether conserved (anti-)BRST charges [cf. Eq. (10)], (ii) the 
appropriate EL-EoM [cf. Eqs. (21),(29)] at suitable places, and (iii) the Gauss divergence theorem at appropriate places.\\


\subsection{Nilpotency Property: Basic (Anti)commutators}

The central theme of our present subsection is to compute the anticommutators: $\{Q_b, \; Q_b \} $ and $\{Q_{ab}, \; Q_{ab} \}$,
respectively, by  using the equal-time {\it basic} canonical (anti)commutators [cf. Eq. (8)] 
of our BRST-quantized theory to prove the nilpotency property
(i.e. $\{Q_b, \; Q_b \} = \frac{1}{2}\, Q_b^2  $ and $\{Q_{ab}, \; Q_{ab} \} = \frac{1}{2}\, Q_{ab}^2  $) of the Noether BRST 
and anti-BRST conserved charges, respectively. In other words, if we can prove that $\{Q_b, \; Q_b \}  = 0$ and $\{Q_{ab}, \; Q_{ab} \} = 0$,
we shall be proving the nilpotency property of the  Noether conserved charges. Toward this goal in our mind, first of all, we note that
the following full form of the anticommutator
\begin{eqnarray}\label{53}
 &&\Big \{ Q_{b}, \; \; Q_{b} \Big \}  = \int \int d^{D-1} x\, d^{D-1} y\,\Big \{ \Big ( 2\, \Pi^0_{(C)}\, \Pi_{(\bar \beta)} - 2\, \Pi^i_{(\bar C)}\, \Pi^{0i}_{(B)} (b) - \Pi^i_{(C)} \, \partial_i  \beta
 \nonumber\\
 &-&  
 \Pi^{ij}_{(B)} (b) \, \big [\partial_i C_j - \partial_j  C_i \big ] - 2\, \Pi^0_{(\bar C)}\, \Pi_{(\phi)} (b) \Big ) (\vec{x},\; t), \; \; 
\Big ( 2\, \Pi^0_{(C)}\, \Pi_{(\bar \beta)} - 2\, \Pi^j_{(\bar C)}\, \Pi^{0j}_{(B)} (b)\nonumber\\
 &-& \Pi^j_{(C)} \, \partial_j  \beta
 - 
 \Pi^{kl}_{(B)} (b) \, \big [\partial_k C_l - \partial_l  C_k \big ] - 2\, \Pi^0_{(\bar C)}\, \Pi_{(\phi)} (b) \Big ) (\vec{y},\; t)
 \Big \},
\end{eqnarray}
is {\it true} if we take into account the expression for the Noether conserved charge $Q_b$ [cf. Eq. (9)]
in terms of the canonical conjugate momenta  [cf. Eq. (7)]
and the basic fields of our BRST-quantized theory (described by the 
{\it coupled} Lagrangian densities ${\cal L}_{(B)}$  and ${\cal L}_{(\bar B)}$ [cf. Eq. (1)]). As per the rules of the 
equal-time {\it basic} canonical (anti)commutators [cf. Eq. (8) and comments after it], we note that the following {\it two} 
non-zero integrals exist from the close and careful observations of the r.h.s. of the above equation, namely;
\begin{eqnarray}\label{54}
 \Big \{ Q_{b}, \; \; Q_{b} \Big \}  &=& \int \int d^{D-1} x\, d^{D-1} y\,\Big \{ \Big (\Pi^i_{(C)} \, \partial_i  \beta \Big ) (\vec{x},\; t), \; \; 
\Big (H^{0kl} \; \partial_k C_l  \Big ) (\vec{y},\; t) \Big \} \nonumber\\
&+& \int \int d^{D-1} x\, d^{D-1} y\,\Big \{  \Big (H^{0kl} \;\partial_k C_l  \Big )  (\vec{x},\; t), \; \; 
\Big (\Pi^i_{(C)} \, \partial_i  \beta \Big ) (\vec{y},\; t) \Big \},
\end{eqnarray}
where we have used the explicit expression for the canonical conjugate momenta: $\Pi^{ij}_{(B)} = \frac{1}{2}\, H^{0ij} $, and
(ii) the totally antisymmetric nature of $H^{0ij}$ which has enabled us to obtain: $\Pi^{kl}_{(B)} (b) \, \big [\partial_k C_l - \partial_l  C_k \big ]
= H^{0kl}\,\partial_k C_l  $. Exploiting the standrad rules of the anticommutators with (i) the composite oprrators, and (ii) the
individual operator (cf. Appendix E), we obtain the explicit form of the r.h.s. of the above equation (54) as follows:
\begin{eqnarray}\label{55}
 \Big \{ Q_{b}, \; \; Q_{b} \Big \}  &=& \int \int d^{D-1} x\, d^{D-1} y\, H^{0kl} (\vec{y},\; t)\,
\Big \{ \Pi^i_{(C)}  (\vec{x},\; t), \; \; 
\big (\partial_k C_l  \big ) (\vec{y},\; t) \Big \} \, \partial_i  \beta (\vec{x},\; t)\nonumber\\
&+& \int \int d^{D-1} x\, d^{D-1} y\, H^{0kl} (\vec{x},\; t)
\Big \{ \big (\partial_k C_l  \big )  (\vec{x},\; t), \; \; 
 \Pi^i_{(C)} (\vec{y},\; t) \Big \}  \partial_i  \beta (\vec{y},\; t).
\end{eqnarray}
At this juncture, we follow a specific set of mathematical exercises. These are  
(i) we use the equal-time basic canonical anticommutator: $\{ C_i (\vec{x},\; t), \; \Pi^j_{(C)} (\vec{y},\; t) \}
= i\, \delta_i^j\, \delta^{D-1} (\vec{x} - \vec{y}) $ [cf. Eq. (8)], (ii) we perform the partial inegration w.r.t. the 
volume integral over $y$, and (iii) we exploit the power of  the Gauss divergence theorem. These exercises lead to 
\begin{eqnarray}\label{56}
 \Big \{ Q_{b}, \; \; Q_{b} \Big \}  &=& - \, 2\, i\, \int \int d^{D-1} x\,\; \Big ( \partial_k H^{0kl} \, \partial_l  \beta
\Big ) (\vec{x},\; t)\;\; \Longrightarrow\;\;\nonumber\\
Q_{b}^2 = \dfrac{1}{2} \Big \{ Q_{b}, \; \; Q_{b} \Big \}\
& = & - \, i\, \int \int d^{D-1} x\; \;\partial_k \,\Big [ H^{0kl} \, \partial_l  \beta
\Big ] (\vec{x},\; t),
\end{eqnarray}
where we have taken into account the property of the Dirac $\delta$-function. Ultimately, we have found that the 
integrand in the above equation turns out to be the total {\it space} derivative. It is due to  the Gauss divergence theorem, at this stage, 
that the above volume integral (56) gets converted into the surface integral where all the physical fields 
tend to go to infinity (and, as a consequence,  vanish off
when $x \to \pm\, \infty$). In such a situation, we note that the Noether conserved BRST charge $Q_b$ becomes 
a BRST invariant (i.e. $s_b Q_b = -\, i\, \{Q_b, \; Q_b \} = 0 $) quantity  as well as nilpotent (i.e.
$Q_b^2 = \frac{1}{2}\, \{Q_b, \; Q_b \} = 0 $) of order two.

Against the backdrop of the above thorough discussions in the context of the explicit computation of the anticommutator
between {\it two} Noether BRST charges (i.e. $\{ Q_b, \; Q_b \} $), we concisely mention a few key steps in the explicit
computation of the anticommutator (i.e. $\{ Q_{ab}, \; Q_{ab} \} $)
between {\it two} Noether conserved anti-BRST charges $Q_{ab}$. In its full blaze of glory, this anticommutator looks as follows
\begin{eqnarray}\label{57}
 &&\Big \{ Q_{ab}, \; \; Q_{ab} \Big \}  = \int \int d^{D-1} x\, d^{D-1} y\,\Big ( 2\, \Pi^0_{(\bar C)}\, \Pi_{(\beta)} 
+ 2\, \Pi^i_{(C)}\, \Pi^{0i}_{(B)} (\bar b) - \Pi^i_{(\bar C)} \, \partial_i \bar \beta \nonumber\\
&-& 
\Pi^{ij}_{(B)} (\bar b) \, \big(\partial_i \bar C_j - \partial_j \bar C_i \big ) - 2\, \Pi^0_{(C)}\, \Pi_{(\phi)} (\bar b) \Big )  (\vec{x},\; t), \; \; 
\Big ( 2\, \Pi^0_{(\bar C)}\, \Pi_{(\beta)} + 2\, \Pi^j_{(C)}\, \Pi^{0j}_{(B)} (\bar b)  \nonumber\\
&-& \Pi^j_{(\bar C)} \, \partial_j \bar \beta
- \Pi^{kl}_{(B)} (\bar b) \, \big(\partial_k \bar C_l - \partial_l \bar C_k \big ) - 2\, \Pi^0_{(C)}\, \Pi_{(\phi)} (\bar b) \Big )  (\vec{y},\; t)
 \Big \},
\end{eqnarray}
where the full form of the expression for $Q_{ab}$ has been taken from equation (9) where we have expressed this Noether anti-BRST charge
in terms of the canonical conjugate momenta [cf. Eq. (7)]  and the basic fields of our BRST-quantized theory described by the 
{\it coupled} Lagrangian densities ${\cal L}_{(B)}$  and ${\cal L}_{(\bar B)}$ [cf. Eq. (1)]). Taking into account the
key features of the equal-time {\it basic} canonical (anti)commutators [cf. Eq. (8)], we note that the following two non-zero integrals
exist from the r.h.s. of the above equation (57). As a consequence, we have the following anticommutator between two 
Noether anti-BRST charges, namely;
\begin{eqnarray}\label{58}
 \Big \{ Q_{ab}, \; \; Q_{ab} \Big \}  &=& \int \int d^{D-1} x\, d^{D-1} y\,\Big \{ \Big (\Pi^i_{(\bar C)} \, \partial_i \bar \beta \Big ) (\vec{x},\; t), \; \; 
\Big (H^{0kl} \; \partial_k \bar C_l  \Big ) (\vec{y},\; t) \Big \} \nonumber\\
&+& \int \int d^{D-1} x\, d^{D-1} y\,\Big \{  \Big (H^{0kl} \;\partial_k \bar C_l  \Big )  (\vec{x},\; t), \; \; 
\Big (\Pi^i_{(\bar C)} \, \partial_i  \bar \beta \Big ) (\vec{y},\; t) \Big \},
\end{eqnarray}
which can be computed, in a straightforward manner, by 
taking into account the rules (cf. Appendix E for more details) of the (anti)commutators  that have been listed in
equation $(E.1)$ . The resulting non-zero integrals are as follows:
\begin{eqnarray}\label{59}
 \Big \{ Q_{ab}, \; \; Q_{ab} \Big \}  &=& \int \int d^{D-1} x\, d^{D-1} y\; H^{0kl} (\vec{y},\; t)\,
\Big \{ \Pi^i_{(\bar C)}  (\vec{x},\; t), \; 
\big (\partial_k \bar C_l  \big ) (\vec{y},\; t) \Big \} \, \partial_i  \bar \beta (\vec{x},\; t)\nonumber\\
&+& \int \int d^{D-1} x\, d^{D-1} y\, H^{0kl} (\vec{x},\; t)
\Big \{ \big (\partial_k \bar C_l  \big )  (\vec{x},\; t), \;  
 \Pi^i_{(\bar C)} (\vec{y},\; t) \Big \}  \partial_i  \bar \beta (\vec{y},\; t).
\end{eqnarray}
At this juncture, we use the equal-time {\it basic} canonical (anti)commutators from equation (8). The outcome 
of this exercise turns out to be the following
\begin{eqnarray}\label{60}
 \Big \{ Q_{ab}, \; \; Q_{ab} \Big \}  &=& - \, 2\, i\, \int \int d^{D-1} x\, \;\Big ( \partial_k H^{0kl} \, \partial_l  \bar \beta
\Big ) (\vec{x},\; t) \;\; \Longrightarrow\;\;\nonumber\\
Q_{ab}^2 = \dfrac{1}{2} \Big \{ Q_{ab}, \; \; Q_{ab} \Big \}\, &=& - \, i\, \int \int d^{D-1} x\; \;
\partial_k \,\Big [ H^{0kl} \, \partial_l  \bar \beta
\Big ] (\vec{x},\; t),
\end{eqnarray}
where we have utilized the property of the Dirac $\delta$-function to obtain the above result. We can apply now the Gauss divergence theorem
because of which the above volume integral gets converted into the  surface integral where 
{\it all} the relevant fields tend to go to infinity. Since
these are physical fields, they vanish off as $x \to \pm \, \infty  $. Hence, we obtain the result that 
the anticommutator (i.e. $\{ Q_{ab}, \; Q_{ab} \} $)
between two Noether conserved anti-BRST charges turns out to be zero (i.e. $\{ Q_{ab}, \; Q_{ab} \} = 0 $). This observation is crucial
in proving that the Noether conserved 
anti-BRST charge $Q_{ab}$ is an anti-BRST invariant (i.e. $s_{ab} Q_{ab} = -\, i\, \{Q_{ab}, \; Q_{ab} \} = 0 $)
quantity and, at the same time, {\it this} conserved charge turns out to be nilpotent (i.e.
$Q_{ab}^2 = \frac{1}{2}\, \{Q_{ab}, \; Q_{ab} \} = 0 $) of order two. \\

\section{ Conclusions}

In our present investigation, we have concentrated on the off-shell nilpotent versions of the (anti-)BRST symmetry transformations [cf. Eq. (2)]
for the D-dimensional BRST-quantized {\it free} Abelian 2-form gauge theory which is endowed with a {\it non-trivial} CF-type
restriction that is responsible for (i) the absolute anticommutativity (i.e. $\{ s_b, \, s_{ab} \} = 0 $)
of the off-shell nilpotent (i.e. $s_{(a)b}^2 = 0 $) (anti-)BRST symmetry transformation operators $s_{(a)b} $ [cf. Eq. $(B.2)$ below], and
(ii) the existence of the coupled (but equivalent) BRST-quantized versions of the Lagrangian densities  [cf. Eq. (1)]. We have
derived (i) the Noether conserved (anti-) BRST currents [cf. Eqs. (3),(4)], and (ii) the Noether conserved 
(anti-)BRST charges [cf. Eqs. (5),(9)]. The {\it latter}
turn out to be the generators for the infinitesimal, continuous and off-shell nilpotent versions of the (anti-)BRST 
transformations (2) which have been proven by taking into account the 
equal-time {\it basic} canonical (anti)commutators [cf. Eq. (8)]
of our D-dimensional BRST-quantized theory (cf. Sec. 2). However, we have explicitly shown that the Noether conserved (anti-)BRST charges
are {\it not} invariant [cf. Eq. (6)]
under the infinitesimal, continuous and {\it off-shell} nilpotent (anti-)BRST symmetry transformations. Thus, the Noether conserved
(anti-)BRST charges are {\it not} the physical quantities w.r.t. the off-shell nilpotent (anti-)BRST
transformations (within the framework of BRST formalism). Hence, they are {\it not} useful in the discussions on the
physicality criteria as far as the basic tenets of the BRST formalism are concerned. We have systematically derived 
the (anti-)BRST {\it invariant} 
versions of the (anti-)BRST charges $Q_{(A)B}$ from their counterpart 
{\it non-invariant} Noether conserved (anti-)BRST charges $Q_{(a)b}$ (see, e.g. [32] for details). In other
words, the conserved (anti-)BRST charges  $Q_{(A)B}$ remain invariant under the off-shell nilpotent (anti-)BRST transformations (2).
It turns out that these {\it invariant} charges, through physicality criteria, lead to the conditions on the physical 
states that are consistent with the Dirac quantization conditions (cf. Sec. 5 for details) for the gauge theories that are endowed with the
{\it first-class} constraints  [in the terminology of Dirac's prescription for the classification scheme 
of constraints (see, e.g. [25-31] for details)].

We have derived the Noether conserved  (anti-)BRST charges [cf. Eq. (5)] 
from the Noether conserved (anti-)BRST currents [cf. Eqs. (3),(4)]. In the systematic derivations  of the (anti-)BRST {\it invariant} charges
$Q_{(A)B}$ from the Noether conserved (anti-)BRST charges $Q_{(a)b}$, we have exploited {\it only} (i) the partial integration along
with the Gauss divergence theorem, and (ii) the appropriate EL-EoM at suitable places
(see, e.g. [32] for details). Hence, it is crystal clear that the {\it modified} versions
of the (anti-)BRST charges $Q_{(A)B}$ are {\it also} conserved quantities. Within the framework of BRST formalism, the conserved and (anti-)BRST
{\it invariant} quantities are the {\it physical} operators (in the true sense of the word)  
which {\it must} be utilized in the physicality criteria (cf. Sec. 5 for details). In contrast to the (anti-)BRST invariant versions of the
(anti-)BRST charges $Q_{(A)B}$, we have shown that the Noether (anti-)BRST charges $Q_{(a)b}$ are {\it not} invariant under the off-shell nilpotent
(anti-)BRST symmetry transformation operators [cf. Eq. (6)]. Hence, to be precise, the Noether conserved (anti-)BRST charges $Q_{(a)b}$ are
{\it not} the physical quantities w.r.t. the off-shell nilpotent (anti-)BRST symmetry transformations (2). This is the reason that the
{\it latter} conserved charges lead to {\it absurd} results when they are utilized in the 
discussions on the physicality criteria within the framework of BRST 
formalism (cf. Sec. 5 for more discussions). Thus, we have been able to show the importance of (i) the Noether conserved (anti-)BRST charges 
$Q_{(a)b}$ as the generators  for the infinitesimal, continuous and {\it off-shell} nilpotent (anti-)BRST transformations (cf. Sec. 2)), 
and (ii) the (anti-)BRST invariant (i.e. $s_{ab} Q_{AB} = 0, \; s_b Q_B = 0 $) versions of the conserved (anti-)BRST charges $Q_{(A)B}$
in the context of the physicality criteria within the framework of BRST formalism (cf. Sec. 5).

We would like to dwell a bit on the existence of the (non-)trivial CF-type restrictions\footnote{The existence of the (non-)trivial 
CF-type restriction(s) is the hallmark of a BRST-quantized gauge invariant theory. These restrictions are deeply
connected with the geometrical object called gerbes. We have corroborated this claim in our earlier works 
on the BRST-quantized versions of the non-Abelian 1-form, free Abelian 2-form and 3-form gauge theories (see, e.g. [37,38] for details).}
on the BRST-quantized version of the gauge/diffeomorphism
invariant theories and the nilpotency property of the Noether (anti-)BRST charges. One of the simplest D-dimensional BRST-quantized version
of a gauge theory is the Abelian 1-form theory where the CF-type restriction is {\it trivial}. As a consequence, we find that the Noether
conserved (anti-)BRST charges are (i) the (anti-)BRST invariant quantities under the {\it off-shell} nilpotent (anti-) BRST symmetry
transformations, (ii) nilpotent of order two, and (iii) the generators for the infinitesimal, continuous and {\it off-shell} nilpotent
(anti-)BRST symmetry transformations. Such nice features are {\it not} found in the case of a D-dimensional BRST-quantized 
non-Abelian 1-form gauge theory (see, e.g. [23] for details)
where the CF-condition [24] is an (anti-)BRST constraint {\it non-trivial} restriction on {\it this} theory. In fact, in our very recent work [23],
we have been able to show that the Noether (anti-)BRST charges are {\it not} invariant under the {\it  off-shell} nilpotent 
(anti-)BRST symmetry transformations {\it even though} these charges are found to be the generators for the off-shell nilpotent transformations.
Hence, these conserved charges are {\it not} physical w.r.t. the off-shell nilpotent (anti-)BRST symmetry transformations. As a consequence, 
these charges lead to {\it absurd} results when they are used in the physicality criteria (see, e.g. Sec 5 of our present work). This is why,
we have consistently {\it modified} the Noether conserved (anti-)BRST charges $Q_{(a)b}$  to the (anti-)BRST invariant
(i.e. $s_{ab} Q_{AB} = 0, \; s_b Q_B = 0 $) versions of the (anti-)BRST charges $Q_{(A)B}$.

As far as the use of the  equal-time {\it basic} canonical (ant)commutators [cf. Eq. (8)] is concerned, we would like to point out that in 
the proof of the  nilpotency property of the Noether conserved (anti-)BRST charges, we have exploited {\it only} the partial integration along with 
the Gauss divergence theorem (cf. Subsec. 6.2 for details). On the contrary, we have found out that, in the proof of the (anti-)BRST
invariance of the {\it modified} versions of the (anti-)BRST charges $Q_{(A)B}$, we have been compelled to use (i) the appropriate EL-EoM
at suitable places, and (ii) the partial integration along with the Gauss divergence theorem (cf. Subsec. 4.2 for details). In exactly similar fashion, we
have observed that {\it both} the above requirements are needed in the proof the nilpotency property of the {\it modified} versions of the
(anti-)BRST charges $Q_{(A)B}$ (cf. Appendix D for details). We have {\it also} proven  the nilpotency property of the Noether 
conserved (anti-)BRST charges
by taking into account the symmetry considerations (cf. Subsec. 6.1 for details). However, in this case, too, we have been compelled to exploit {\it both}
the above conditions. Thus, we conclude that the proof of the nilpotency property of the Noether conserved (anti-)BRST charges by using the
equal-time {\it basic} canonical (anti-)commutators 
(cf. Subsec. 6.2 for details) is superior to the {\it same} proof by using the symmetry considerations (cf. Subsec. 6.1 for details).

We have devoted quite a number of years in the study of the higher $p$-form ($p = 2, 3$) form gauge theories (see, e.g. [11-13],[15],[16]
and references therein) where the (anti-)BRST invariant CF-type restrictions are {\it non-trivial}. It would be nice future endeavor to apply
the ideas of (i) our earlier work [23] on the D-dimensional non-Abelian 1-form gauge theory, and (ii) our present work, to the D-dimensional
BRST-quantized version of the Abelian 3-form gauge theories to check the nilpotency property and (anti-)BRST invariance of the Noether conserved
(anti-)BRST charges. Furthermore, we envisage to study the physicality criteria, within the framework of BRST formalism, for (i) the {\it massless}
Abelian 3-form gauge theory, and (ii) the St{\" u}ckelberg-modified {\it massive} Abelian 3-form gauge theory, in any arbitrary dimension
of spacetime. These are the future directions we plan to pursue for our investigation(s) and our results will be reported in 
our forthcoming publications.\\

\vskip 0.5 cm

\begin{center}
{\bf Appendix A: On the Gauge-Fixing and Faddeev-Popov Ghost Terms}\\
\end{center}

\vskip 0.3 cm

\noindent
The purpose of our present Appendix is to express the gauge-fixing and FP-ghost terms of equation (1) in terms of the
off-shell nilpotent (i.e. $s_{(a)b}^2 = 0 $) (anti-)BRST symmetry transformation operators $s_{(a)b} $ [cf. Eq. (2)].
In this connection, we note the following\footnote{All the numerical factors (along with their precise signs) are very important
as far as the {\it coupled} nature of the Lagrangian densities 
${\cal L}_{(B)} $ and ${\cal L}_{(\bar B)} $ is concerned.
This statement is made in connection with {\it all} the individual terms that are located inside the square bracket (cf. Appendix B, too, for more
discussions).}
\[
{\cal L}_{(B)}  = \dfrac{1}{12} \, H^{\mu\nu\sigma} \, H_{\mu\nu\sigma}
+ s_b \, s_{ab} \, \Big [\dfrac{1}{4}\, \big (B_{\mu\nu} \big )^2 - \dfrac{1}{2} \, \bar C^\mu C_\mu  + \dfrac{1}{4}\,\big (\phi \big )^2  
+ \dfrac{1}{4}\, \bar \beta\, \beta \Big ], 
\]
\[
{\cal L}_{(\bar B)}  = \dfrac{1}{12} \, H^{\mu\nu\sigma} \, H_{\mu\nu\sigma}
- s_{ab} \, s_b \, \Big [\dfrac{1}{4}\, \big (B_{\mu\nu} \big )^2 - \dfrac{1}{2} \, \bar C^\mu C_\mu  + \dfrac{1}{4}\,\big (\phi \big )^2 
+ \dfrac{1}{4}\, \bar \beta\, \beta \Big ],
\eqno(A.1)
\]
where all the terms, located within the square brackets, carry (i) the mass dimension (D-2) in the natural units $\hbar = c = 1 $ because
the applications of the (anti-)BRST symmetry transformation operators $s_{(a)b} $ raise the mass dimension 
of a specific field by one  on which they act [cf. Eq. (2)], and (ii) the ghost number equal to zero because every term 
of the final version
of the BRST-quantized Lagrangian density must have the ghost number equal to {\it zero} which is ensured by
the applications of $s_b$ and $s_{ab}$ {\it together} in equation (A.1).

The observations in the above equation (A.1) 
 prove, in a straightforward manner, the {\it perfect} BRST and anti-BRST invariance of the coupled Lagrangian densities 
${\cal L}_{(B)} $ and ${\cal L}_{(\bar B)} $, respectively, because of the fact that (i) the kinetic term: $\frac{1}{12}\, H^{\mu\nu\sigma}\, H_{\mu\nu\sigma} $
remains invariant under the BRST 
and anti-BRST symmetry transformations, and (ii) the BRST (i.e. $s_b $) and anti-BRST (i.e. $s_{ab} $) symmetry transformation
operators $s_{(a)b} $
are found to be off-shell nilpotent (i.e. $s_{(a)b}^2 = 0 $) of order two. To be precise, we would like to add that the explicit computations
of the {\it top} entry, in the above equation (A.1), leads to the expression for the Lagrangian density 
${\cal L}_{(B)}$ [cf. Eq. (1)]
{\it plus} the total spacetime derivative term as: $-\, \partial_\mu [B^{\mu\nu}\, B_\nu + \frac{1}{2}\, \rho \, C^\mu +  \frac{1}{2}\, \lambda \, 
\bar C^\mu]$.  On the other hand, we obtain ${\cal L}_{(\bar B)}$ [cf. Eq. (1)] 
{\it plus} the total spacetime derivative term: $ \partial_\mu [B^{\mu\nu}\, \bar B_\nu - \frac{1}{2}\, \rho \, C^\mu -  \frac{1}{2}\, \lambda \, 
\bar C^\mu]$ from the explicit compuation of the {\it lower} entry 
 from the above equation (A.1). Since the total spacetime derivatives terms, added to any arbitrary Lagrangian density, 
do {\it not} affect the dynamics of the theory (which is described by {\it that} Lagrangian density), it is crystal clear that we have obtained the {\it coupled} 
Lagrangian densities ${\cal L}_{(\bar B)} $ and ${\cal L}_{(B)} $ from the explicit computation of the equation (A.1) where {\it these}
Lagrangian densities have been expressed as the  {\it sum} of (i) the kinetic term, and (ii) the additional terms that are written in terms
of the (anti-)BRST symmetry transformation operators $s_{(a)b} $ acting on the specific combination of fields.\\

\vskip 0.5 cm 
\begin{center}
{\bf Appendix B: Coupled Lagrangian Densities, CF-Type Restriction and Off-Shell Nilpotent Symmetry Considerations}\\
\end{center}

\vskip 0.3 cm 

\noindent
The main objective of our present Appendix is to prove the existence of the (anti-)BRST invariant CF-type restriction:
$B_\mu + \bar B_\mu + \partial_\mu \phi = 0 $  on our D-dimensional BRST-quantized version of the free Abelian 2-form 
gauge theory from different theoretical angels.  First of all, we note that the EL-EoM w.r.t. the {\it bosonic}
 Nakanishi-Lautrup auxiliary fields 
$B_\mu$ and  $B_\mu$, from the {\it coupled} Lagrangian densities ${\cal L}_{(B)} $ and ${\cal L}_{(\bar B)} $,  respectively, are as follows:
\[
B_\mu = \partial^\nu B_{\nu\mu} - \dfrac{1}{2}\, \partial_\mu \phi, \;\;\qquad \;\;
\bar B_\mu = -\, \partial^\nu B_{\nu\mu} - \dfrac{1}{2}\, \partial_\mu \phi.
\eqno(B.1)
\] 
It is straightforward to note that the CF-type restriction:  $B_\mu + \bar B_\mu + \partial_\mu \phi = 0 $
ensue from the above EL-EoM. However, this observation is {\it not} a formal proof of
the derivation of the  CF-type restriction. Rather, it justifies the {\it correct} coupled nature of
the Lagrangian  densities ${\cal L}_{(B)}$ and 
${\cal L}_{(\bar B)}$ [cf. Eq. (1)]. Furthermore, we observe that the absolute
anticommutativity property (i.e $\{ s_b, \; s_{ab} \} = 0$) of the (anti-)BRST symmetry
transformation operators ($s_{(a)b} $) is found to be {\it true} for {\it all} the fields of our 
D-dimensional BRST-quantized field-theoretic system of the free 
Abelian 2-form gauge theory {\it except} the following
\[
\big \{s_b, \; s_{ab} \big \} \, B_{\mu\nu} = - \partial_\mu \big (B_\nu + \bar B_\nu \big )
+ \partial_\nu \big (B_\mu + \bar B_\mu \big ), 
\eqno(B.2)
\] 
which {\it also} turns out be {\it zero} if we invoke the validity of the CF-type restriction: $B_\mu + \bar B_\mu + \partial_\mu \phi = 0 $
on our theory which is {\it physical} in the
sense that it is an (anti-)BRST invariant (i.e. $ s_{(a)b} [B_\mu + \bar B_\mu + \partial_\mu \phi] = 0$)
quantity [cf. Eq. (2)]. The requirements of the nilpotency
and absolute anticommutativity properties are the {\it defining} features of the infinitesimal and 
continuous (anti-)BRST symmetry transformation operators.
We note, in  our present discussion, that if we demand the absolute anticommutativity property: $\{s_b, \; s_{ab} \} = 0 $
between the BRST and anti-BRST transformation operators, we are compelled to have 
the existence of the CF-type restriction: $B_\mu + \bar B_\mu + \partial_\mu \phi = 0 $ on our BRST-quantized theory.

We are in the position now to concentrate on the  {\it equivalence} of the {\it coupled} Lagrangian  densities ${\cal L}_{(B)} $ and ${\cal L}_{(\bar B)} $ 
w.r.t. the off-shell nilpotent (anti-)BRST symmetry transformations. In this context, we note that {\it both} the Lagrangian densities
respect the BRST as well as the anti-BRST symmetry transformations {\it together} provided the whole theory is considered on the subspace (of the 
entire space) of quantum fields where the CF-type restriction (i.e. $B_\mu + \bar B_\mu + \partial_\mu \phi = 0 $) is
satisfied. To corroborate this statement, we note that, in addition to
our observations in equation (3), we have the following transformations:
\[
s_{ab} {\cal L}_{(B)} = -\, \partial_\mu \Big [(\partial^\mu \bar C^\nu - \partial^\nu \bar C^\mu)\, B_\nu + (\partial_\nu B^{\nu\mu} +
\dfrac{1}{2}\, \bar B^\mu )\, \rho - \dfrac{1}{2}\, \lambda\, \partial^\mu \bar \beta \Big ]
\]
\[
+ (\partial^\mu \bar C^\nu - \partial^\nu \bar C^\mu)\, \partial_\mu [B_\nu + \bar B_\nu + \partial_\nu \phi]
+ \dfrac{1}{2}\,[B_\mu + \bar B_\mu + \partial_\mu \phi]\, \partial^\mu \rho,
\]
\[
s_{b} {\cal L}_{(\bar B)} = +\, \partial_\mu \Big [(\partial^\mu C^\nu - \partial^\nu  C^\mu)\, \bar B_\nu + (\partial_\nu B^{\nu\mu} -
\dfrac{1}{2}\, B^\mu )\, \lambda + \dfrac{1}{2}\, \rho\, \partial^\mu \beta \Big ]
\]
\[
- (\partial^\mu  C^\nu - \partial^\nu  C^\mu)\, \partial_\mu [B_\nu + \bar B_\nu + \partial_\nu \phi]
+ \dfrac{1}{2}\,[B_\mu + \bar B_\mu + \partial_\mu \phi]\, \partial^\mu \lambda.
\eqno(B.3)
\]
The above transformations become total spacetime derivatives 
\[
s_{ab} {\cal L}_{(B)} = -\, \partial_\mu \Big [(\partial^\mu \bar C^\nu - \partial^\nu \bar C^\mu)\, B_\nu + (\partial_\nu B^{\nu\mu} +
\dfrac{1}{2}\, \bar B^\mu )\, \rho - \dfrac{1}{2}\, \lambda\, \partial^\mu \bar \beta \Big ]
\]
\[
s_{b} {\cal L}_{(\bar B)} = +\, \partial_\mu \Big [(\partial^\mu C^\nu - \partial^\nu  C^\mu)\, \bar B_\nu + (\partial_\nu B^{\nu\mu} -
\dfrac{1}{2}\, B^\mu )\, \lambda + \dfrac{1}{2}\, \rho\, \partial^\mu \beta \Big ]
\eqno(B.4)
\]
if we invoke the validity of the CF-type restriction:
$B_\mu + \bar B_\mu + \partial_\mu \phi = 0 $. In other words, the
{\it coupled} Lagrangian  densities ${\cal L}_{(B)} $ and ${\cal L}_{(\bar B)} $ are {\it equivalent} from the point of view of the symmetry
considerations on the subspace (of the {\it total} space) of the quantum fields where the CF-type restriction is satisfied. In fact,
on this subspace, we observe that (i) {\it both} the Lagrangian densities respect the (anti-)BRST symmetry transformations
{\it together}, and (ii) these symmetry operators absolutely anticommute (i.e. $\{ s_b, \; s_{ab} \} = 0 $) 
with each-other thereby maintaining their own independent identity.
The observations in (B.3) are, in some sense, the formal derivations of the CF-type restriction (from the point
of view of the requirement that {\it both} the Lagrangian densities ${\cal L}_{(B)} $ and ${\cal L}_{(\bar B)} $ {\it must} respect the off-shell nilpotent
versions of the 
(anti-)BRST transformations to be called as {\it equivalent} in the true sense of the word within the framework of BRST formalism).

We end this Appendix with the {\it final} remark that {\it coupled} Lagrangian  densities ${\cal L}_{(B)} $ and ${\cal L}_{(\bar B)} $ are 
{\it truly} equivalent (on the subspace of the quantum fields where the CF-type restriction: $B_\mu + \bar B_\mu + \partial_\mu \phi = 0 $ is satisfied)
provided we prove {\it it} from the requirement that: ${\cal L}_{(B)} - {\cal L}_{(\bar B)} = 0 $. In fact, {\it this} requirement leads to the 
following observation 
\[
{\cal L}_{(B)} - {\cal L}_{(\bar B)} = \Big [\partial^\nu B_{\nu\mu} + \dfrac{1}{2} \, \big (\bar B^\mu - B^\mu \big ) \Big ]\;
\Big (B_\mu + \bar B_\mu + \partial_\mu \phi \Big ) = 0,
\eqno(B.5)
\]
where we have used (i) the simple trick of the 
factorization: $(\bar B^\mu \, \bar B_\mu - B^\mu \,  B_\mu ) = (\bar B^\mu + B^\mu)\,(\bar B_\mu - B_\mu) $,
and (ii) the straightforward result that: $ (\partial_\nu B^{\nu\mu})\,\partial_\mu \phi = \partial_\mu (\partial_\nu B^{\nu\mu}\, \phi)$.
Hence, in this short (but quite important) Appendix, we have been able to prove (i) the coupled (but equivalent) nature of the Lagrangian  
densities ${\cal L}_{(B)} $ and ${\cal L}_{(\bar B)} $, and (ii) the formal proof of the
existence of the (anti-)BRST invariant CF-type restriction
(i.e. $B_\mu + \bar B_\mu + \partial_\mu \phi = 0 $)
on our D-dimensional BRST-quantized version of the free Abelian 2-form gauge theory\footnote{It is worthwhile to mention that, in our earlier work [39], 
we have derived (i) the off-shell nilpotent (anti-)BRST symmetry transformations (2), and (ii) the (anti-)BRST invariant CF-type restriction
by applying the superfield approach to the D-dimensional free Abelian 2-form gauge theory.}.\\

\vskip 0.5 cm 
\begin{center}
{\bf Appendix C: On the First-Class Constraints}\\
\end{center}

\vskip 0.3 cm

\noindent
It is a well-known fact that the gauge theories are {\it always} endowed with the first-class constraints in the terminology of 
Dirac's classification scheme for constraints (see, e.g. [25-31]). Our starting Lagrangian density [i.e. ${\cal L}_{(0)} = \frac{1}{12} \, H^{\mu\nu\sigma}\, H_{\mu\nu\sigma} $]
for the free D-dimensional Abelian 2-form [i.e. $B^{(2)} =  \frac{1}{2!}\, B_{\mu\nu}\, (dx^\mu \wedge dx^\nu) $] gauge field $B_{\mu\nu}$ is
also endowed with a set of {\it two} first-class constraints.  To corroborate this claim, let us begin with
\[
{\cal L}_{(0)} = \dfrac{1}{12}\, H^{\mu\sigma}\, H_{\mu\nu\sigma}, \qquad \;\;
H^{(3)} = d \, B^{(2)} = \dfrac{1}{3!}\, H_{\mu\nu\sigma}\, (d x^\mu \wedge dx^\nu\wedge d x^\sigma),
\eqno(C.1)
\]
where (i) the totally antisymmetric field-strength tensor: $H_{\mu\nu\sigma} = \partial_\mu B_{\nu\sigma} + \partial_\nu B_{\sigma\mu} 
+ \partial_\sigma B_{\mu\nu} $ [with $\mu, \nu, \sigma...= 0, 1, 2.....(D-1) $] has been derived from the Abelian 3-form 
(i.e $H^{(3)} = d\, B^{(2)}] $), and (ii) the exterior derivative: $ d = \partial_\mu (dx^\mu)$ (with $d^2 = 0$) raises the degree
of a form by one (on which it operates). It is straightforward to note that the canonical conjugate momenta (i.e. $\Pi^{\mu\nu}_{(B)} $)
w.r.t. the gauge field $B_{\mu\nu}$ are as follows
\[
\Pi_{(B)}^{\mu\nu}  = \dfrac{\partial \, {\cal L}_{(0)}}{\partial\, (\partial_0\, B_{\mu\nu})} = \dfrac{1}{2}\, H^{0\mu\nu}
 \; \;\; \Longrightarrow \;\;\;  \Pi_{(B)}^{0i} = \dfrac{1}{2}\, H^{00i} \approx 0, \qquad \Pi_{(B)}^{ij} = \dfrac{1}{2}\, H^{0ij},
\eqno(C.2)
\]
where (i) the component  $\Pi_{(B)}^{0i} \approx 0 $, of the above canonical conjugate momenta $\Pi_{(B)}^{\mu\nu} $,
is the primary constraint (PC) on our theory, and (ii) the mathematical symbol $\approx 0 $ stands for 
the technical word {\it weakly} zero in the context of the classification scheme of constraints (which physically implies
that we are allowed to take a first-order time derivative on it).

To obtain the secondary constraint (SC) on our theory, we have to demand the time-evolution invariance of the PC (where we are allowed to take
the first-order time derivative on the PC). It is well-known observation that the Hamiltonian formalism is the most suitable approach to 
determine the full set of constraints on a  theory.  However, for our simple D-dimensional {\it free} Abelian 2-form gauge theory, the EL-EoM
(derived from the Lagrangian density of our theory)  can yield the SC [30]. Toward this goal in mind, we note that the following EL-EoM,
derived from the Lagrangian density (C.1), namely;
\[
\dfrac{1}{2}\, \partial_\mu H^{\mu\nu\sigma} = 0 \; \;\; \Longrightarrow \;\;\;
 \dfrac{1}{2}\,    \Big (\partial_0 H^{00j} + \partial_i H^{i0j} \big ) = 0,
\eqno(C.3)
\]
where we have (i) chosen the indices: $\nu = 0, \; \sigma = j $, and (ii) summed over the repeated index $\mu$. It is obvious from equation (C.3) 
that we have obtained the requirement of the time-evolution invariance of the PC  (i.e. $\Pi_{(B)}^{0i} \approx 0 $) as follows:
\[
 \partial_0  \Big [ \dfrac{1}{2}\,H^{00j} \Big ] =  \partial_i \Big [ \dfrac{1}{2}\,H^{0ij} \Big ] \approx 0 
\; \;\; \Longrightarrow \;\;\; \partial_0 \Pi^{0j}_{(B)} = \partial_i \Pi^{ij}_{(B)} \approx 0.
\eqno(C.4)
\]
As a consequence, we have already obtained the SC as $\partial_i \Pi^{ij}_{(B)} \approx 0 $ (which has emerged out, we re-emphasize,  from the 
requirement of the time-evolution
invariance of the PC). For our present simple system, many studies have been 
performed and it has been found that there are {\it no} further constraints on our theory.
Since we find that the PC (i.e. $\Pi^{0i}_{(B)} \approx 0 $)  and the SC (i.e. $\partial_i \Pi^{ij}_{(B)} \approx 0$) are expressed in terms of the 
components of the canonical conjugate momenta $\Pi_{(B)}^{\mu\nu} $, {\it these} constraints will commute\footnote{To be 
precise, at the {\it classical} level, the Poisson bracket between two constraints will be zero. However, we have used the 
word ``commute" in view of the fact that, ultimately, our free Abelian 2-form gauge theory will be BRST-quantized where the 
Poisson brackets will get converted into the commutators.}
 with each-other (as per the rules
of the canonical quantization scheme). Hence, we conclude that our D-dimensional free Abelian 2-form gauge theory is endowed with 
a set of {\it two} first-class constraints
in the terminology of Dirac's prescription for the classification scheme of constraints.\\

\vskip 0.5 cm 
\begin{center}
{\bf Appendix D: On the Nilpotency Property of $Q_{(A)B}$}\\
\end{center}

\vskip 0.3 cm

\noindent
The central purpose of our present short Appendix is to prove, using the equal-time {\it basic} canonical (anti)commutators of our equation (8), 
 that the (anti-)BRST invariant (i.e. $s_{ab} Q_{AB} = 0, \; s_b Q_B = 0 $)
versions of the (anti-)BRST charges $Q_{(A)B}$ (consistently derived from the Noether conserved charges $Q_{(a)b}$) 
are {\it not} nilpotent (i.e. $Q_{(A)B}^2 \neq 0 $) of order two if we do {\it not} use (i) the partial integration along with
the Gauss divergence theorem, and (ii) the appropriate EL-EoM at suitable places in our proof. To corroborate the above claim, first of all, 
let us focus on the study of the
nilpotency property of the modified version of the BRST charge $Q_B$ by explicitly computing the
anticommutator: $\{Q_B, \; Q_B \} $. It is straightforward to note that, in its full blaze of glory, this anticommutator
looks as 
\[
 \Big \{ Q_{B}, \; \; Q_{B} \Big \}  = \int \int d^{D-1} x\, d^{D-1} y\,\Big \{  \big [\partial^0 B^i - \partial^i B^0 \big ]\, C_i +
\Pi^i_{(C)}\, \partial_i  \beta - 2\, \Pi^i_{(\bar C)}  \,  \Pi^{0i}_{(B)} (b)
\]
\[
- \dot \rho\, \beta
- 2\, \Pi^0_{(C)} \, \Pi_{(\bar \beta)}  + 2\, \Pi^0_{(\bar C)} \, \Pi_{(\phi)} (b) \Big ) (\vec{x},\; t), \; \; \;
\Big ( \big [\partial^0 B^j - \partial^j B^0 \big ]\, C_j +
\Pi^j_{(C)}\, \partial_j  \beta 
\]
\[- 2\, \Pi^i_{(\bar C)}  \,  \Pi^{0i}_{(B)} (b)
- \dot \rho\, \beta
- 2\, \Pi^0_{(C)} \, \Pi_{(\bar \beta)}  + 2\, \Pi^0_{(\bar C)} \, \Pi_{(\phi)} (b)  \Big ) (\vec{y},\; t)
 \Big \},
\eqno(D.1)
\]
where the expression for the {\it modified} version of the BRST charge $Q_B$, in terms of the 
canonical conjugate momenta, has been taken from equation (45).
Out of the above full form of the terms that are present on the r.h.s. of
 the anticommutator $\{ Q_{B}, \;  Q_{B}  \} $, the actual non-zero existing anticommutators
(for our purpose) are as follows:
\[
 \Big \{ Q_{B}, \; \; Q_{B} \Big \}  = \int \int d^{D-1} x\, d^{D-1} y\,\Big \{  \big [\partial^0 B^i - \partial^i B^0 \big ]\, C_i\Big )
(\vec{x},\; t), \; \; \Big (
\Pi^j_{(C)}\, \partial_j  \beta \Big ) (\vec{y},\; t) \Big \}
\]
\[
+\, \int \int d^{D-1} x\, d^{D-1} y\,\Big \{ \Big (
\Pi^i_{(C)}\, \partial_i  \beta \Big ) (\vec{x},\; t), \; \; \Big (\big [\partial^0 B^j - \partial^j B^0 \big ]\, C_j \Big ) (\vec{y},\; t) \Big \}.
\eqno(D.2)
\]
Using the standard rules of the (anti)commutators involving the composite and/or the individual operators
(see, e.g.  Appendix E), we end-up with the following explicit non-zero anticommutators
in view of the rules laid down by the canonical brackets (8), namely;
\[
 \Big \{ Q_{B}, \; \; Q_{B} \Big \}  = \int \int d^{D-1} x\, d^{D-1} y\, \big (\big [\partial^0 B^i - \partial^i B^0 \big ] \big) (\vec{x},\; t) \;
 \Big \{ C_i (\vec{x},\; t), \; \; \Pi^j_{(C)} (\vec{y},\; t) \Big \}\,  \partial_j  \beta (\vec{y},\; t)
 \] 
 \[
+\, \int \int d^{D-1} x\, d^{D-1} y\,\big (\big [\partial^0 B^j - \partial^j B^0 \big ] \big) (\vec{y},\; t) \;
  \Big \{ \Pi^i_{(C)} (\vec{x},\; t), \; \; C_j (\vec{y},\; t) \Big \}\,  \partial_i  \beta (\vec{x},\; t).
\eqno(D.3)
\]
At this crucial juncture, we use (i) the basic equal-time
canonical anticommutator between the ghost fields and their canonical conjugate momenta:
 $\{ C_i (\vec{x},\; t), \; \; \Pi^j_{(C)} (\vec{y},\; t) \} = i\,
\delta_i^j \, \delta^{(D - 1)} (\vec{x} - \vec{y} ) $ [cf. Eq. (8)], and (ii) the standard rules of the Direc $\delta$-function in performing
the partial integration over the volume integral corresponding to the $y$ variable. This
exercise leads us to  obtain the {\it final} result as follows
\[
 Q_B^2 = \dfrac{1}{2}\, \Big \{ Q_{B}, \; \; Q_{B} \Big \} \equiv i\, \int d^{D-1} x\, \Big ( \big [\partial^0 B^i - \partial^i B^0 \big ]  
\;  \partial_i  \beta \Big ) (\vec{x},\; t) \neq 0,
\eqno(D.4)
\]
which demonstrates that the {\it direct} computation of the anticommutator $\{ Q_{B}, \;  Q_{B}  \} $ is found to be {\it non-zero}.
In other words, the {\it modified} version of the BRST charge $Q_B$ is found to be non-nilpotent (i.e. $Q_B^2 \neq 0$) by
our direct computation of $\{ Q_{B}, \;  Q_{B}  \} $. However, if we use the EL-EoM:
$\partial_i H^{0ij} = (\partial^o B^j - \partial^j B^0) $ [cf. Eq. (21)] which is derived form the Lagrangian density ${\cal L}_{(B)}$
[cf. Eq. (1)] w.r.t. the gauge field $B_{\mu\nu}$, we find that the above integral [on the r.h.s.
of equation $(D.4)$] takes the following interesting form:
\[
 Q_B^2 = \dfrac{1}{2}\, \Big \{ Q_{B}, \; \; Q_{B} \Big \}  = i\, \int d^{D-1} x\, \Big ( \big [\partial_i H^{0ij} ]  
\;  \partial_j  \beta \Big ) (\vec{x},\; t) 
\]
\[
\equiv  i\, \int d^{D-1} x\, \Big (\partial_i \big [ H^{0ij} \, \partial_j \beta \big ] \Big )
(\vec{x},\; t) \; \;\;\Longrightarrow \; \; \; 0.
\eqno(D.5)
\]
In other words, the integrand on the r.h.s. of equation $(D.4)$ becomes a total {\it space} derivative term. At this stage.
if we utilize  the theoretical strength of 
the Gauss divergence theorem, we observe that the anticommutator $\{ Q_{B}, \;  Q_{B}  \} $ becomes {\it zero}
(because all the physical fields 
vanish off as $x \to \pm\, \infty $).  As a consequence, we 
conclude  that the
{\it modified} version of the BRST charge $Q_B$ becomes nilpotent (i.e. $Q_B^2 = 0 $) of order two.

Against the backdrop of the above detailed discussions, we very {\it concisely} discuss the explicit computation of the anticommutator:
$\{ Q_{AB}, \;  Q_{AB}  \} $ to shed light on the nilpotency property of the {\it modified} version of the anti-BRST charge $Q_{AB}$. First
of all, we note that the full form of the anticommutator $\{ Q_{AB}, \;  Q_{AB}  \} $ is as follows
\[
 \Big \{ Q_{AB}, \; \; Q_{AB} \Big \}  = \int \int d^{D-1} x\, d^{D-1} y\,\Big \{ \Big (\Pi^i_{(\bar C)}\, \partial_i  \bar \beta
- \big [\partial^0 \bar B^i - \partial^i \bar B^0 \big ]\, \bar C_i + 2\, \Pi^i_{(C)}  \,  \Pi^{0i}_{(B)} (\bar b)
\]
\[
- \dot \lambda\, \bar \beta
+ 2\, \Pi^0_{(\bar C)} \, \Pi_{(\beta)}  - 2\, \Pi^0_{(C)} \, \Pi_{(\phi)} (\bar b) \Big ) (\vec{x},\; t), \; \; \;
\Big (\Pi^i_{(\bar C)}\, \partial_i  \bar \beta
- \big [\partial^0 \bar B^i - \partial^i \bar B^0 \big ]\, \bar C_i + 2\, \Pi^i_{(C)}  \,  \Pi^{0i}_{(B)} (\bar b)
\]
\[
- \dot \lambda\, \bar \beta
+ 2\, \Pi^0_{(\bar C)} \, \Pi_{(\beta)}  - 2\, \Pi^0_{(C)} \, \Pi_{(\phi)} (\bar b)    \Big ) (\vec{y},\; t)
 \Big \},
\eqno(D.6)
\]
where we have used the expression for $Q_{AB}$ in terms of the canonical conjugate momenta from equation (37).
Taking into account the equal-time {\it basic} canonical (anti)commutators from equation (8), we find that the non-zero existing anticommutators
(of our interest) are: 
\[
 \Big \{ Q_{AB}, \; \; Q_{AB} \Big \}  = - \,\int \int d^{D-1} x\, d^{D-1} y\,
\Big \{ \Big (\big [\partial^0 \bar B^i - \partial^i \bar B^0 \big ]\, \bar C_i\Big )
(\vec{x},\; t), \; \; \Big (
\Pi^j_{(\bar C)}\, \partial_j  \bar \beta \Big ) (\vec{y},\; t) \Big \}
\]
\[
-\, \int \int d^{D-1} x\, d^{D-1} y\,\Big \{ \Big (
\Pi^i_{(\bar C)}\, \partial_i  \bar \beta \Big ) (\vec{x},\; t), \; \; 
\Big (\big [\partial^0 \bar B^j - \partial^j \bar B^0 \big ]\, \bar C_j \Big ) (\vec{y},\; t) \Big \}.
\eqno(D.7)
\]
Both the above integrals, on the r.h.s. of the above equation, are equal. To corroborate this statement, we find that the above
equation $(D.7)$ can be explicitly re-written as:
\[
 \Big \{ Q_{AB}, \,  Q_{AB} \Big \}  = - \int \int d^{D-1} x\, d^{D-1} y\, \big (\big [\partial^0 \bar B^i 
- \partial^i \bar B^0 \big ] \big) (\vec{x}, t) \;
 \Big \{ \bar C_i (\vec{x}, t), \, \Pi^j_{(\bar C)} (\vec{y}, t) \Big \}\,  \partial_j \bar  \beta (\vec{y}, t)
 \] 
 \[
-\, \int \int d^{D-1} x\, d^{D-1} y\,\big (\big [\partial^0 \bar B^j - \partial^j \bar B^0 \big ] \big) (\vec{y},\; t) \;
  \Big \{ \Pi^i_{(\bar C)} (\vec{x},\; t), \; \;\bar  C_j (\vec{y},\; t) \Big \}\,  \partial_i  \bar \beta (\vec{x},\; t).
\eqno(D.8)
\]
Exploiting the beauty and theoretical strength of the equal-time {\it basic} canonical anticommutator (i.e. $\{\bar C_i (\vec{x}, t), \; \Pi^j_{(\bar C)} \vec{y}, t)\} = i\, \delta_i^j\,
\delta^{(D-1)} (\vec{x} - \vec{y}) $)
between the anti-ghost fields $\bar C_i$ and the corresponding canonical conjugate momenta
$\Pi^i_{(\bar C)}$, we obtain the following explicit result from the above equation $(D.8)$, namely;
\[
 Q_{AB}^2 = \dfrac{1}{2}\, \Big \{ Q_{AB}, \; \; Q_{AB} \Big \} \equiv -\, i\, \int d^{D-1} x\, 
\Big ( \big [\partial^0 \bar B^i - \partial^i \bar B^0 \big ]  
\;  \partial_i  \bar \beta \Big ) (\vec{x},\; t) \neq 0,
\eqno(D.9)
\]
which shows that the {\it modified} version of the anti-BRST charge $Q_{AB}$ is found to be non-nilpotent
(i.e. $Q_{AB}^2 \neq 0 $) from the direct computation of the anticommutator $\{ Q_{AB}, \;  Q_{AB}  \} $. However, if we use the
EL-EoM that has been quoted in equation (29), at this juncture, we observe that the following is {\it true}, namely;
\[
 Q_{AB}^2 = \dfrac{1}{2}\, \Big \{ Q_{AB}, \; \; Q_{AB} \Big \}  = +\, i\, \int d^{D-1} x\, \Big ( \big [\partial_i H^{0ij} ]  
\;  \partial_j  \bar \beta \Big ) (\vec{x},\; t) 
\]
\[
\equiv +\,  i\, \int d^{D-1} x\, \Big (\partial_i \big [ H^{0ij} \, \partial_j \bar \beta \big ] \Big )
(\vec{x},\; t) \; \;\;\Longrightarrow \; \; \; 0.
\eqno(D.10)
\]
In other words, we find that the integrand of equation $(D.9)$ becomes a total {\it space} derivative term when we use the EL-EoM
w.r.t. the gauge field $B_{\mu\nu}$ [cf. Eq. (29)]
that is derived from the Lagrangian density ${\cal L}_{(\bar B)}$ [cf. Eq. (1)]. As far as equation $(D.10)$ is concerned,
we note that we have applied the theoretical strength of the Gauss divergence theorem which implies that the
anticommutator $\{ Q_{AB}, \;  Q_{AB}  \} $ is equal to {\it zero} (i.e. $\{ Q_{AB}, \;  Q_{AB}  \} = 0 $) because
all the physical fields vanish off as $x \to \pm \, \infty $. This proves the nilpotency ($Q_{AB}^2 = 0$) of $Q_{AB}$.

We end this Appendix with a couple of concluding remarks. First of all, we note that (i) the Noether conserved (anti-)BRST charges $Q_{(a)b}$ are 
{\it not} invariant [cf. Eq. (6)] under the off-shell nilpotent (anti-)BRST symmetry transformations (2), and (ii) the {\it modified}
versions of the (anti-)BRST charges $Q_{(A)B}$  {\it not} nilpotent [cf. Eqs. $(D.4), (D.9)$] if we use {\it only} the
equal-time {\it basic} canonical brackets form equation (8) in the verification of the standard relationships: $Q_B^2 = \frac{1}{2}\,\{Q_B, \; Q_B \}$ and
$Q_{AB}^2 = \frac{1}{2}\,\{Q_{AB}, \; Q_{AB} \}$ through the explicit computation of the anticommutators $\{ Q_{B}, \;  Q_{B}  \}$
and $\{ Q_{AB}, \;  Q_{AB}  \}$. Second, if we use (i) the partial integration along with the Gauss divergence theorem, and (ii) the appropriate EL-EoM 
at suitable places, we find that (i) the Noether conserved (anti-)BRST charges become (anti-)BRST invariant as well as nilpotent of
order two (cf. Sec. 6 for more discussions), and (ii) the {\it modified} versions of the (anti-)BRST charges $Q_{(A)B}$ {\it also} 
become (anti-)BRST invariant as well as nilpotent of order two\footnote{it is crystal clear that the (anti-)BRST invariant  
(i.e. $s_{ab} Q_{AB} = 0, \, s_b Q_B = 0$) versions of 
the {\it physical} (anti-)BRST charges $Q_{(A)B}$ will be useful in the discussions on the BRST cohomology (see, e.g. [30]).}. 
These observations should be contrasted against 
the backdrop of our earlier work  on the D-dimensional BRST-quantized non-Abelian 1-form gauge theory [23]
where we have shown that the {\it modified} versions of the (anti-)BRST charges $Q_{(A)B}$ are found to be 
non-nilpotent (i.e. $Q^2_{(A)B} \neq 0 $). \\

\vskip 0.5 cm 
\begin{center}
{\bf Appendix E: On the Standard Rules for (Anti)commutators}\\
\end{center}

\vskip 0.3 cm

\noindent
The decisive feature of our present Appendix is to collect a few {\it basic} (anti)commutator rules that have been used in the explicit computations
of some of the key equations in Subsecs. 4.2 and 6.2. In this context, we use
the following {\it standard} rules of the (anti)commutators 
for the composite as well as the independent operator(s) at various stages 
of our theoretical discussions. These very useful relationships amongst the bosonic and fermionic 
(composite as well as independent) operators are as follows
\[
\big  \{ {\cal F}_1\, {\cal B}_1, \; {\cal F}_2 \big \} = {\cal F}_1 \, \big  [ {\cal B}_1, \; {\cal F}_2 \big ] 
+ \big  \{ {\cal F}_1, \; {\cal F}_2 \big \} \, {\cal B}_1, 
\]
\[
\big  \{ {\cal B}_1\, {\cal F}_1, \; {\cal F}_2 \big \} = {\cal B}_1 \, \big  \{ {\cal F}_1, \; {\cal F}_2 \big \} 
- \big  [{\cal B}_1, \; {\cal F}_2 \big ] \, {\cal F}_1, 
\]
\[
\big  \{ {\cal F}_1, \; {\cal B}_2\, {\cal F}_2 \big \} =  \big  [ {\cal F}_1, \; {\cal B}_2 \big ] \, {\cal F}_2 
+ {\cal B}_2 \, \big  \{ {\cal F}_1, \; {\cal F}_2 \big \}, 
\]
\[
\big  \{{\cal F}_1, \; {\cal F}_2 {\cal B}_2 \big \} = \big  \{ {\cal F}_1, \; {\cal F}_2 \big \}\, {\cal B}_2 
- {\cal F}_2\,  \big  [{\cal F}_1, \;{\cal B}_2 \big ] , 
\]
\[
\big  [{\cal F}_1\,{\cal F}_2, \;{\cal F}_3 \big ] ={\cal F}_1 \, \big  \{ {\cal F}_2, \; {\cal F}_3 \big \} 
- \big  \{ {\cal F}_1, \; {\cal F}_3 \big \} \, {\cal F}_2, 
\]
\[
\big  [{\cal F}_1, \;{\cal F}_2\, {\cal F}_3 \big ] =  \big  \{ {\cal F}_1, \; {\cal F}_2 \big \} \, {\cal F}_3 
- {\cal F}_2\, \big  \{{\cal F}_1, \;{\cal F}_3 \big \},
\eqno(E.1)
\]
where a couple of operators ${\cal B}_1$ and ${\cal B}_2$ are bosonic 
(i.e. ${\cal B}_1 {\cal B}_2 = {\cal B}_2 {\cal B}_1, \; {\cal B}_1^2 \neq 0,   {\cal B}_2^2 \neq 0$)
in nature. On the other hand, a set of {\it three} operators ${\cal F}_1$, ${\cal F}_2$ and ${\cal F}_3$ are fermionic 
(i.e. ${\cal F}_1 {\cal F}_2 = -\,{\cal F}_2 {\cal F}_1, \, {\cal F}_1 {\cal F}_3 = -\, {\cal F}_3{\cal F}_1, 
\, {\cal F}_2 {\cal F}_3 = -\, {\cal F}_3 {\cal F}_2, \; {\cal F}_1^2 = 0, \,  {\cal F}_2^2 = 0, \,  {\cal F}_3^2 = 0 $) in nature.

We end this very short Appendix with the remark that even though we have listed the above (anti)commutators for {\it only} two bosonic 
operators (i.e. ${\cal B}_1, \, {\cal B}_2 $) and three fermionic operators (i.e. ${\cal F}_1, \, {\cal F}_2, \, {\cal F}_3 $), the algebraic relationships
$(E.1)$ can be used for the computations of the (anti)commutators for multiple bosonic and fermionic operators. A close and careful look at
equation $(E.1)$ shows that we have taken only {\it three} operators on the l.h.s. However, these algebraic relationships of
the above equation $(E.1)$ can be used when we shall be dealing with a set of $N$ number (i.e. $N > 3 $)  of operators by
recasting them in the forms of $(E.1)$ with {\it judicious}
 choices of the composite operators in a specific manner. In other words, 
we can always assume a set of operators as {\it one} operator to recast any arbitrary (anti)commutator of the $N$-number 
(i.e. $N > 3 $) of operators in the form that is similar to $(E.1)$. This process can be repeated depending on the number of
operators (i.e. $N > 3 $) to {\it finally} obtain the given (anti)commutator {\it exactly} in the form of  $(E.1)$ at the end
of all the possible assumptions/choices of the operators.\\

\end{document}